# Analyzing and generating multimode optical fields using self-configuring networks

David A. B. Miller

*Ginzton Laboratory, Stanford University, 348 Via Pueblo Mall, Stanford CA 94305 USA*
*dabm@stanford.edu*

Working with finite numbers of modes to describe, generate and detect optical fields can be both mathematically economical and physically useful. Such a modal basis can map directly to various applications in communications, sensing and processing. But, we need a way to generate and analyze such fields, including measurement and control of both the relative amplitudes and phases of the modal components. Here, we show first how to measure all those relative amplitudes and phases automatically and simultaneously. The method uses a self-configuring network of 2x2 blocks, such as integrated Mach-Zehnder interferometers, that can automatically align itself to the optical field by a sequence of simple one-parameter power minimizations when network elements, such as phase shifters, are adjusted. The optical field is then directly deduced from the resulting settings of those elements. We show how the entire network can be calibrated for such measurements, automatically and with just two light beams. Then, using the same calibration and running the mesh backwards, we can also controllably generate an arbitrary multimode field. Explicit algorithms and formulas are given for operating this system.

## 1. Introduction

Increasingly in optics, we exploit complex multimode light fields [1] – for example, in communications [2], classical and quantum linear optical information processing [3-5], and in various opportunities in adaptive systems for sensing and communications [6]. To extract the complete optical information from the multimode fields, we need to analyze the relative amplitude and phase of the different modes. Similarly, we want to be able to generate arbitrary multimode field inputs for linear computation and communication or for full control of an optical stimulus in sensing. Schemes based on single spatial light modulators (SLMs) have shown impressive performance in sequentially analyzing such multimode fields (e.g., [7,8]); a single SLM can also generate an arbitrary field, though generally with some inherent power loss. Other multiplane schemes based on several SLMs or fixed diffractive elements can simultaneously separate components of different specific modes from large multimode fields (e.g., [9,10]). Though such separations can give the relative magnitudes of each mode, a full analysis also requires their relative phases, which requires further interferometry. With some further measurements and processing in those schemes, full-field interferometry can analyze the full complex field [7]. Here, however, we show how self-configuring integrated meshes of waveguide interferometers (e.g., Mach-Zehnder interferometers (MZIs)) can automatically measure and generate both the relative amplitudes and phases of such multimode fields, potentially even tracking changing inputs in real time [11,12]. Furthermore, we show how such a mesh can be calibrated automatically for these purposes, using just two simple light beams.

Our approach exploits the ability of certain classes of mesh architectures to self-configure, without calculations, to align to the input multimode field [13,14]. Then, the resulting values of the various parameters in the mesh – e.g., phase shifts and/or coupling ratios – contain all the necessary information about the input field. Essentially, in this self-configuration, we have performed all the relative interferometric measurements across all the input modes. We can then deduce the relative amplitudes and phases in the input field from these mesh parameters using only simple arithmetic. Furthermore, by running such a mesh system backwards using a single source light beam injected "backwards" into the "output", any such multimode field can instead be controllably generated, emerging from the original "inputs".

The mesh itself operates on the field in its multiple different input (single-mode) waveguides. Such meshes could operate directly on a "free-space" optical input field by sampling it with grating couplers [6,11,13-15]; alternatively, they could exploit multiplane [9,10] or other mode separation techniques like a photonic lantern [16] as a fixed "front end" preprocessor that transforms from large overlapping continuous spatial modes to the multiple discrete waveguide mesh inputs, allowing use of the technique with an arbitrary choice of modal basis.

Many sophisticated mesh networks of interferometers have been demonstrated that can work efficiently with complex multimode optical fields to provide linear operations [3,4,11,15,17-30]. Several architectures offer self-configuration [6,11,13-15,31-34], with the simplest being a self-aligning beam coupler architecture [13], which is at the core of this paper; this can take an arbitrary set of (mutually-coherent) inputs in a set of waveguides and deliver all the power to just one output waveguide. This self-configuration is based on a progressive set of simple one-parameter feedback loops to set the values of the various phase shifters and/or coupling elements in the mesh based on detected powers, and without any calculations. In the automatic calibration process, we can also "calibrate out" the phase behavior of any "front-end" optics used to interface the mesh to the external optical world. A key point in this approach is that both the required calibration and the self-configuration and measurement could be completely automated using simple progressive algorithms and control circuits followed by elementary arithmetic calculations.

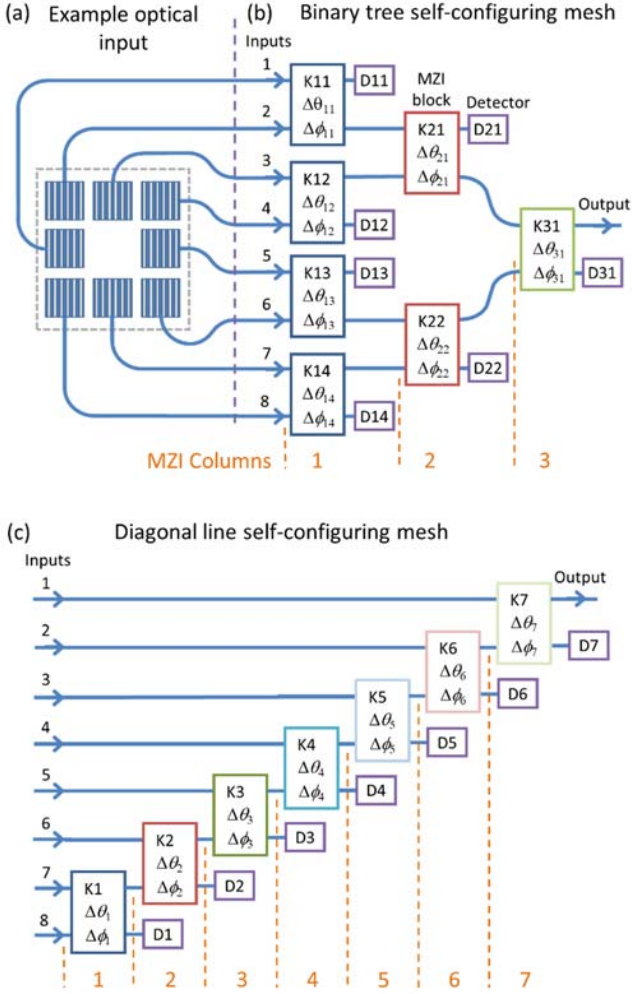

Fig. 1. Concept and architectures of a self-configuring mesh layer with optical inputs. (a) Example optical input, here with 8 waveguides driven by the outputs from an example square array of grating couplers illuminated by some input light field. (b) Binary tree mesh, which we can self-configure to give all the resulting input power in the one Output waveguide. Parameters $\Delta\theta$ and $\Delta\phi$ control the "split ratio" and one other phase shift in each 2x2 block K, respectively. Each such block can be implemented as an MZI. Elements D are detectors at the "drop ports" of these blocks or MZIs. (c) Alternative, diagonal line mesh architecture. The successive "columns" of MZIs are shown for both architectures.

Though any optical component may have some losses due to imperfections, and any such waveguide systems are likely to have input coupling losses, otherwise these mesh analysis and generation schemes are perfectly efficient optically; there are no other power losses in the system, even for arbitrary (mutually coherent) inputs and/or outputs in the single-mode input or output waveguides in the meshes. Recent analysis suggests that such self-alignment can be accomplished in microseconds or shorter even when working with optical input powers of only ~ 10 microwatts per input [12] (given sufficiently fast adjustable phase shifters and/or couplers in the mesh). Hence such an approach could be convenient, optically efficient, and fast.

In this paper, we discuss the approach of using self-configuring architectures for analyzing multimode beams in Section 2. In Section 3, we summarize the necessary general description of an MZI block. Section 4 shows how to deduce the input field from the self-configured mesh settings. Section 5 shows how to use the mesh as an arbitrary multimode generator. Calibration processes are summarized in Section 6. We draw conclusions in Section 7. Supporting detail is given in the Supplement.

## 2. Automatic analysis of a multimode beam

### Self-configuring architectures

Two basic forms of self-configuring architectures of interferometers allow the self-aligning beam coupler [13] – those based on a "diagonal line" and those using a "binary tree" (see Fig. 1). Hybrids of these are also possible. (For completeness, the topology of such architectures is discussed in the Supplement Section S1, together with some other alternative architectures.) Each of the "diagonal line" and "binary tree" mesh architectures as shown here can be a "layer" of a general self-configuring architecture (the Supplement Section S1). Though we work here with only one such layer, multiple such layers can be cascaded for more complex functions, including arbitrary linear processors [14]. Each such layer consists of multiple "columns" of MZIs [35]. For the diagonal line, there is only one MZI per column, though in the binary tree there may be several in a given column.

For a "stand-alone" analyzer/generator, the binary tree may be preferable because (1) it allows some parallelization of the analysis process, (2) it is shorter – for $N$ input or output waveguides, it requires only $\sim \log_2 N$ "columns" rather than the $\sim N-1$ "columns" of a diagonal line, and (3) all paths from the inputs to the output go through the same number of MZI, and hence can have the same background loss. (Generally, background loss does not affect the functionality of the architectures and meshes here as long as it is equal on all paths; then it just results in some uniform overall loss in the system. As a result, we will analyze our systems here as if they are lossless, with the understanding that there may be some such overall loss factor.) The "diagonal line" approach may be preferable if we want to cascade multiple self-configuring layers (it can then lead to shorter architectures, without any crossing waveguides), and has the additional feature of being essentially symmetric from "front" to "back".

For mutually coherent light at the inputs, these architectures can self-configure to direct all the input power in the various input waveguides to the one output waveguide. This self-configuration can be achieved using a succession of single-parameter power minimizations at each of the "drop-port" detectors D11, D12, … or D1, D2, … . The self-configuring algorithms can be remarkably simple [13,14,32,34]. With the light field of interest shining on the optical inputs, the relative phase at the input to an MZI is adjusted first to minimize the "drop-port" power, and then the "split ratio" of the MZI is adjusted to take that power to zero, and the MZIs are set this way in sequence starting with the MZI(s) in Column 1, and proceeding through those in Column 2, and so on [13]. Multiple MZIs in a given Column (as in the binary tree) can all be configured at the same time, in parallel (see [35] and the Supplement Section S1). It is also possible to self-configure the mesh based on power maximization just using a detector at the output (e.g., D31 or D7), though without the simple parallelization of the process in that case [13].

**Analysis algorithm and modal basis**

The basic algorithm for using such a self-configuring layer to analyze a multimode beam is shown in Fig. 2(a). Once we have calibrated the mesh (see Section 6), we can shine in the optical field of interest, self-configure the mesh to route all the power to the output, and then deduce relative amplitudes and phases of the different parts or modes of the original beam from the resulting settings of the mesh. (The corresponding process for generating a desired output "backwards" from the inputs is shown in Fig. 2(b) and is discussed in Section 5.)

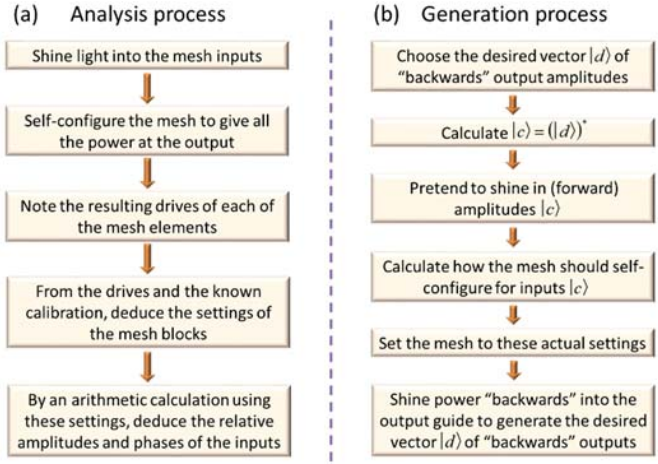

Fig. 2. Algorithm outlines (a) for automatic analysis of a multimode input field, (b) generating a desired "backwards" multimode output field.

Used directly with grating couplers as in Fig. 1(a), the "native" modes of this system corresponding to separate beams on the different grating couplers, each with the phase corresponding to the phase of the phase reference beam used in the system calibration (see Section 6) at the corresponding coupler; the modal analysis is done first on this basis.

If instead we use some fixed front-end mode transformer such as multiplane converters [9,10] or photonic lanterns [16] that deliver various different external (and likely overlapping) modes to different waveguide outputs that directly feed the inputs of the mesh, then those external modes become the effective modal basis on which this analysis is performed. Since the analyzer here has performed all the necessary physical interferometry, it is also possible in a simple mathematical calculation on those measured amplitude and phase values to transform the results after measurement to some other basis; essentially, we multiply the measured output vector of (complex) amplitudes by an appropriate unitary matrix $U_B$ to change the modal basis for the measurement. That then means that the effective modal basis for the measurement is made up from specific different (and orthogonal) linear combinations (or input "vectors") of the input signals; indeed, the rows of the matrix $U_B$ are then just the Hermitian adjoints (i.e., conjugate transpose) of those orthogonal input vectors. Hence, with this additional simple mathematical calculation (a matrix multiplication), this approach can be used to measure on any modal basis that is supported by the optics.

Now we need to understand how the settings of an MZI relate to the input amplitudes and phases. So, we formally analyze the MZI (Section 3), then deduce the expressions relating the settings to these field parameters (Section 4).

## 3. Analysis of Mach-Zehnder interferometer

Here we need a full form of the analysis of an MZI. We summarize this here, with necessary notations, and with supporting detail in the Supplement Section S2.

Though only two phase shifters are required to give sufficient degrees of freedom in an MZI for it to function as a universal 2x2 block, for a general and flexible notation, we analyze it as if it has four phase shifters, as in Fig. 3(a). In use, we only need at least one phase shifter on at least one arm inside the MZI, and any other one of these four phase shifters. We label the four ports of the MZI analogously to the four ports of a conceptual "cube" beamsplitter (Fig. 3(b)), as "Top" (T), "Left" (L), "Right" (R) and "Bottom" (B) – so, with the notion that light can enter at the "Top" port and be partially reflected to the "Right" port, as well as partially transmitted to the "Bottom" port, and similarly light incident in the "Left" port will be partially reflected to the "Bottom" port and partly transmitted to the "Right" port. We also label the arms inside the interferometer as "Upper" (P) and "Lower" (W). 50:50 waveguide beamsplitters BS1 and BS2, shown as directional couplers, couple light between the upper and lower waveguides, to complete the MZI.

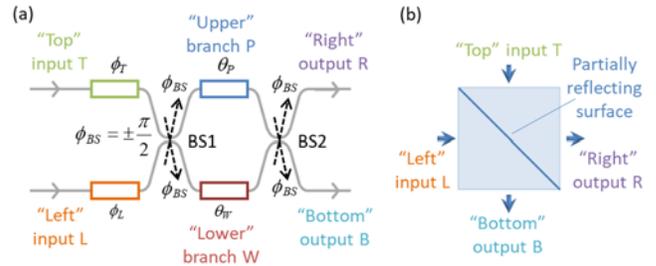

Fig. 3. (a) Schematic of a general waveguide MZI with up to four phase shifters (colored rectangles) on waveguides (grey lines), with two 50:50 beamsplitters BS1 and BS2, shown as directional couplers. (b) Corresponding conceptual beamsplitter showing "Top", "Left", "Bottom" and "Right" surfaces.

We need to define various phase delays, as shown in Fig. 3. We take $\phi_T$ to be the total phase delay from the "Top" input T, through the upper waveguide to the left end of the "Upper" branch phase shifter, marked with $\theta_P$. Similarly $\phi_L$ is the total phase delay through the lower waveguide from the "Left" input L to the left end of the "Lower" branch phase shifter marked with $\theta_W$. $\theta_P$ and $\theta_W$ are similarly the total phase delays on the upper and lower waveguides, respectively, from the left ends of these "Upper" and "Lower" phase shifters, respectively, to the "Right" and "Bottom" output ports, respectively.

Note there is necessarily an additional phase shift $\phi_{BS}$, for light passing "through" the beamsplitters from one waveguide to the other. (Our convention on phase shifts here is that a positive phase shift to corresponds to a phase delay.) For a lossless beamsplitter (whether or not it has a 50:50 split), this phase shift is necessarily either $+\pi/2$ or $-\pi/2$ (within arbitrary additive multiples of $2\pi$); this follows from power conservation (see the Supplement Section S3) and hence mathematical unitarity in a lossless system [14,36]. Whether the sign is "+" or "–" depends on the specifics of the beamsplitter (see the Supplement Section S4). For a directional coupler beamsplitter in a "first order" design (i.e., with the shortest length for a given coupling to the "other" guide), this is a phase delay $+\pi/2$ (i.e., for a beam starting in one guide, the beam coupled over into the "other" guide is delayed by 90° compared to the beam remaining in the original guide). For definiteness now, and because this is the most likely form of beam splitter anyway, we choose $\phi_{BS} = +\pi/2$. (Note, incidentally, that for any such "shortest" directional coupler beamsplitter, even if it does not have a 50:50 beamsplitter ratio, this phase shift is still $+\pi/2$. Only if we lengthen the coupler so much that it moves into a "second order" behavior – coupling "over" and partially "back" again – does this number change to $-\pi/2$, or equivalently $+3\pi/2$.)

Though we will be physically controlling phase shifts such as $\phi_T$, $\phi_L$, $\theta_P$ and/or $\theta_W$ – for example, a common approach is to control just the $\phi_T$ and $\theta_P$ phase shifters – the description of the MZI factorizes most simply using other resulting phases. For the two interferometer arms, we can usefully define the "common mode"

$$\theta_{av} = \frac{\theta_P + \theta_W}{2} \quad (1)$$

and "differential"

$$\Delta\theta = \theta_P - \theta_W \quad (2)$$

phase shifts. Similarly, for the input arms we have "common mode"

$$\phi_{av} = \frac{\phi_T + \phi_L}{2} \quad (3)$$

and "differential"

$$\Delta\phi = \phi_T - \phi_L \quad (4)$$

phase shifts. We can also usefully define an overall phase shift from these

$$\phi_{Tot} = \phi_{av} + \theta_{av} + \frac{\pi}{2} \quad (5)$$

(the inclusion of the additional $\pi/2$ makes later algebra simpler).

We can use the interference within an MZI to allow it to calibrate itself, in which case we calibrate these differential phases $\Delta\theta$ and $\Delta\phi$ (see Section 6) as a function of drive. We also have to know how we created these differential phase shifts – for example, by driving only the $\phi_T$ and $\theta_P$ phase shifters – so we can calculate the corresponding change in $\phi_{Tot}$ from Eqs. (1), (3), and (5) that is associated with any chosen setting of $\Delta\theta$ and $\Delta\phi$.

With input (complex) amplitudes $a_T$ and $a_L$ respectively for the propagating modes in the "Top" and "Left" input waveguides, leading to resulting propagating mode amplitudes $a_R$ and $a_B$ of the beams exiting in the "Right" and "Bottom" output waveguides, as derived in the Supplement Section S2, we can write the relation between inputs and outputs generally as

$$\begin{bmatrix} a_R \\ a_B \end{bmatrix} = \mathsf{M}(\phi_{Tot}, \Delta\theta, \Delta\phi) \begin{bmatrix} a_T \\ a_L \end{bmatrix} \quad (6)$$

Here we are regarding the 2x2 matrix $\mathsf{M}(\phi_{Tot}, \Delta\theta, \Delta\phi)$ for a given MZI as a function of $\phi_{Tot}$, $\Delta\theta$ and $\Delta\phi$, which are themselves each functions of some of $\phi_T$, $\phi_L$, $\theta_P$ and/or $\theta_W$. This choice of matrix parameters also allows us factorize $\mathsf{M}$ as

$$\mathsf{M}(\phi_{Tot}, \Delta\theta, \Delta\phi) = \exp(i\phi_{Tot}) \mathsf{M}_s(\Delta\theta) \mathsf{M}_\phi(\Delta\phi) \quad (7)$$

with

$$\mathsf{M}_\phi(\Delta\phi) = \begin{bmatrix} \exp\left(i\frac{\Delta\phi}{2}\right) & 0 \\ 0 & \exp\left(-i\frac{\Delta\phi}{2}\right) \end{bmatrix} \quad (8)$$

For the simplest (and desired) case of 50:50 beamsplitters we obtain

$$\mathsf{M}_s(\Delta\theta) = \begin{bmatrix} \sin\left(\frac{\Delta\theta}{2}\right) & \cos\left(\frac{\Delta\theta}{2}\right) \\ \cos\left(\frac{\Delta\theta}{2}\right) & -\sin\left(\frac{\Delta\theta}{2}\right) \end{bmatrix} \quad (9)$$

The full form of $\mathsf{M}_\phi(\Delta\phi)$ for other beamsplitter ratios is given in Eq. (S7) in the Supplement Section S2.

For use of the mesh as a generator, we need to run it "backwards" (i.e., shining light backwards into the "Right" and/or "Bottom" ports to have light emerge from the "Top" and/or "Left" ports). Then by Eq. (S58) of the Supplement Section S5, the resulting matrix $\mathsf{B}$ in this direction is just the transpose of $\mathsf{M}$ (i.e., $\mathsf{M}^T$). (We also want this backwards matrix $\mathsf{B}$ when performing the calculations to analyze an input field.) Given the specific factorization of Eq. (7) and these matrix symmetries, then the "backwards" matrix is

$$\begin{aligned} \mathsf{B}(\phi_{Tot}, \Delta\theta, \Delta\phi) &= \mathsf{M}^T(\phi_{Tot}, \Delta\theta, \Delta\phi) \\ &= \exp(i\phi_{Tot}) \mathsf{M}_\phi^T(\Delta\phi) \mathsf{M}_s^T(\Delta\theta) \end{aligned} \quad (10)$$

where we note that the order of the matrix product in Eq. (10) has been inverted compared to that of Eq. (7) as a consequence of the transpose of the matrix product inverting the product order. (Matrix $\mathsf{M}_\phi$ (Eq. (8)) is symmetric, so it is its own transpose, but we keep the transpose notation for similarity with later Hermitian adjoint algebra. Though $\mathsf{M}_s$ (Eq. (9)) is also symmetric for this ideal 50:50 beamsplitter case, in general it is not (see Eq. (S7) in the Supplement Section S2), so we retain the explicit transpose for it also.)

So with amplitudes $b_R$ and $b_B$ of backwards waves entering the "Right" and "Bottom" ports, respectively, and $b_T$ and $b_L$ emerging from the "Top" and "Left" ports respectively, we have formally

$$\begin{bmatrix} b_T \\ b_L \end{bmatrix} = \mathsf{B}(\phi_{Tot}, \Delta\theta, \Delta\phi) \begin{bmatrix} b_R \\ b_B \end{bmatrix} \quad (11)$$

with $\mathsf{B}$ as in Eq. (10).

## 4. Deducing the input vector from the Mach-Zehnder settings

Suppose, then, that we had shone the input vector $|c\rangle = \begin{bmatrix} c_1 & c_2 & \ldots & c_N \end{bmatrix}^T$ of complex amplitudes into the $N$ input waveguides on the "left" in Fig. 1. (Here and below we can use the Dirac notation $|c\rangle$ as a shorthand for such a column vector of complex amplitudes.) Our approach to analyzing the relative amplitudes and phases of all these input forward amplitudes, Fig. 2. Algorithm outlines (a) for automatic analysis of a multimode input field, (b) generating a desired "backwards" multimode output field.(a), involves having the mesh self-configure so that, for such an input vector, all of the power appears in the single "Output" waveguide (at the "Right" output of the MZI in the last column of the self-configuring layer, as in Fig. 1(b) or (c)). Then, from the settings we now have in the MZIs, we wish to deduce the input vector $|c\rangle$ that led to these settings.

One simple way to envisage this calculation of $|c\rangle$ from the resulting mesh settings is to imagine that, with the mesh set this way, we now run the mesh backwards, shining (unit amplitude) light backwards into this Output waveguide. Then we calculate the vector $|d\rangle \equiv [d_1 \quad d_2 \quad ... \quad d_N]^T$ that would emerge "backwards" from the input guides. Then, by the phase conjugating property of unitary meshes run backwards (the Supplement Section S5), and noting that a backwards-propagating beam in a single-mode guide is just the phase conjugate of a forwards-propagating beam in that guide, we deduce

$$|c\rangle \equiv [c_1 \quad c_2 \quad ... \quad c_N]^T = ([d_1 \quad d_2 \quad ... \quad d_N]^T)^* \equiv (|d\rangle)^* \tag{12}$$

completing the analysis of the input beam relative amplitudes and phases. In other words, the input vector $|c\rangle$ of amplitudes we calculate this way must have been the input vector of amplitudes that self-configured the mesh to these settings.

So, for each MZI we can note the drives that we are applying to the phase shifters as a result of the self-configuration process. Hence using our presumed known calibration of the mesh elements, we can deduce the corresponding settings $\Delta\theta$ and $\Delta\phi$ that the self-configuration has set for this MZI. Again, knowing how we had driven the mesh phase shifters (i.e., what specific ones or combinations of $\phi_T$, $\phi_L$, $\theta_P$ and/or $\theta_W$ we actually drove in calibration and in self-configuration and what drives we had applied), we can also now calculate the change in $\phi_{Tot}$ from Eqs. (1), (3), and (5). (We will already have compensated for any fixed phases implicit in $\phi_{Tot}$ as part of our phase calibration process, so such a change is all we need to know – see the discussion of calibration below in Section 6.) Hence, we can calculate the corresponding matrix $\mathsf{M}$ as in Eq. (7) for each MZI, or, more usefully here, the corresponding "backwards" matrix $\mathsf{B}$ as in Eq. (10).

We can always construct a full $N \times N$ matrix for any mesh with $N$ inputs and $N$ outputs by progressively multiplying together the various 2x2 matrices for each block. Doing this appropriately for the backwards matrices $\mathsf{B}$, we could calculate the corresponding $N \times N$ backwards matrix for the mesh, and we could perform calculations with that matrix. For our situation, though, we are only interested in calculating the backwards field at the "input" ports for hypothetical light in just one "output" port. As a consequence, we can set up the result even more simply and directly. We can write for each MZI

$$\mathsf{B} = \begin{bmatrix} \alpha & \mu \\ \beta & \nu \end{bmatrix} \tag{13}$$

where we know all of these elements $\alpha$, $\beta$, $\mu$, and $\nu$ from the calculation using Eq. (10) for each block. Then we start with a hypothetical backwards "input" vector in the block in the last column (e.g., block K31 or K7 in Fig. 1) with $b_R = 1$ and $b_B = 0$, (a vector $[1 \quad 0]^T$) corresponding to shining hypothetical unit field backwards into the appropriate Output waveguide in Fig. 1. We can then simply work progressively "backwards" to the input, allowing us to write explicit expressions for each input waveguide amplitude. For this discussion, we subscript the $\alpha$, $\beta$, $\mu$, and $\nu$ in each block with the corresponding block number.

For example, for the binary tree mesh in Fig. 1(b), working backwards from the output, the backwards output vector at the left ports of K31 is $[\alpha_{31} \quad \beta_{31}]^T$, so the backwards amplitude in the top left port of K31 is $\alpha_{31}$. That then feeds the bottom right port of block K21, giving a backwards "input" vector at the right of block K21 of $[0 \quad \alpha_{31}]^T$, so the output backwards amplitude at the top left port of block K21 is $\mu_{21}\alpha_{31}$. Continuing similarly backwards through block K11, the top left output amplitude is $\mu_{11}\mu_{21}\alpha_{31}$. So, proceeding similarly for the various other backwards paths, we have explicitly for an 8-input mesh as in Fig. 1

$$|d\rangle \equiv \begin{bmatrix} d_1 \\ d_2 \\ d_3 \\ d_4 \\ d_5 \\ d_6 \\ d_7 \\ d_8 \end{bmatrix} = \begin{bmatrix} \mu_{11}\mu_{21}\alpha_{31} \\ \nu_{11}\mu_{21}\alpha_{31} \\ \alpha_{12}\nu_{21}\alpha_{31} \\ \beta_{12}\nu_{21}\alpha_{31} \\ \mu_{13}\alpha_{22}\beta_{31} \\ \nu_{13}\alpha_{22}\beta_{31} \\ \alpha_{14}\beta_{22}\beta_{31} \\ \beta_{14}\beta_{22}\beta_{31} \end{bmatrix} \tag{14}$$

So, explicitly again for such an 8 input binary tree mesh, we would conclude that the relative amplitudes and phases of the input beams in the 8 input waveguides as used for the self-configuration must have been

$$|c\rangle \equiv \begin{bmatrix} c_1 \\ c_2 \\ c_3 \\ c_4 \\ c_5 \\ c_6 \\ c_7 \\ c_8 \end{bmatrix} = \begin{bmatrix} (\mu_{11}\mu_{21}\alpha_{31})^* \\ (\nu_{11}\mu_{21}\alpha_{31})^* \\ (\alpha_{12}\nu_{21}\alpha_{31})^* \\ (\beta_{12}\nu_{21}\alpha_{31})^* \\ (\mu_{13}\alpha_{22}\beta_{31})^* \\ (\nu_{13}\alpha_{22}\beta_{31})^* \\ (\alpha_{14}\beta_{22}\beta_{31})^* \\ (\beta_{14}\beta_{22}\beta_{31})^* \end{bmatrix} \equiv (|d\rangle)^* \tag{15}$$

formally completing the analysis of the multimode input field. This process is easily extended for larger meshes (e.g., 16x1, 32x1, etc.). So, from the knowledge of how each MZI has been set in the self-configuration process with input amplitudes as in $|c\rangle$, we can calculate all the $\alpha$, $\beta$, $\mu$, and $\nu$ matrix elements for each block, and hence deduce from Eq. (15) what the original input vector of field mode amplitudes $|c\rangle$ – the one used to self-configure the mesh – actually was.

Note that this approach works for any of the choices of the pair of phase shifters (i.e., two phase shifters out of the 4, of which at least one is on an MZI internal arm). (If we use only the phase shifters on the MZI internal arms, we need additional phase shifters on at least one input to each MZI in the "input" column (Column 3 in the example of Fig. 1), and the common-mode phase shifts in MZIs in one column end up performing the equivalent of the functions of the $\phi_T$ and $\phi_L$ phase shifters in the next column to the right.) For the 8 input diagonal line mesh as in Fig. 1(c), the corresponding result for $|c\rangle$ is, similarly,

$$|c\rangle = \begin{bmatrix} (\alpha_7)^* \\ (\alpha_6\beta_7)^* \\ (\alpha_5\beta_6\beta_7)^* \\ (\alpha_4\beta_5\beta_6\beta_7)^* \\ (\alpha_3\beta_4\beta_5\beta_6\beta_7)^* \\ (\alpha_2\beta_3\beta_4\beta_5\beta_6\beta_7)^* \\ (\alpha_1\beta_2\beta_3\beta_4\beta_5\beta_6\beta_7)^* \\ (\beta_1\beta_2\beta_3\beta_4\beta_5\beta_6\beta_7)^* \end{bmatrix} \qquad (16)$$

## 5. Running the mesh as a generator

With a calibrated mesh, it is also straightforward to run the mesh in reverse, starting with actual light power fed backwards into the Output port as in Fig. 1, and using the mesh as a generator of any specific vector $|d\rangle$ of complex amplitudes emerging backwards from the "inputs" of the mesh. The algorithm is summarized briefly in Fig. 2(b).

To understand how to set the various drives in the mesh to generate some such "backwards" output vector, we can imagine we are configuring the mesh to take a vector $|c\rangle \equiv (|d\rangle)^*$ being shone into the mesh inputs in the forward direction to hypothetically put all the power in the output waveguide. If we set the mesh as if it was "collecting" this forward vector $|c\rangle$, then when run in the backwards direction, it will generate the desired vector $|d\rangle$ of outputs from the "input" ports. We give explicit details for the calculations of the required mesh parameters in the Supplement Section S6.

Note that, if we are using the mesh as a generator when operating with some additional "front-end" optics, such as such as multiplane converters [9,10] or photonic lanterns [16], it is important that, in that external optics, the loss is the same for all modes; otherwise, the whole system is not unitary within some overall loss factor, and the phase-conjugating property of unitary networks run backwards does not apply. Then running the whole system backwards as a generator would not lead to the desired backwards output; it would not be the phase-conjugate of the hypothetical forward vector $|c\rangle$ of modal amplitudes at the front end of the entire optical system.

## 6. Calibration

When working with complex meshes in applications such as these, it is essential that the mesh and its elements are calibrated. There are two aspects to the calibration: (i) "split ratio" (or $\Delta\theta$) calibration of the MZIs; and (ii) phase (or $\Delta\phi$) calibration. It is obviously useful that such calibration is simple and automatic.

In calibrating the phase in the system, we need to choose an external phase reference of some kind. That could be just some flat phase front shining directly into the waveguides of the integrated optics; but, more generally, it could be a phase reference such as a plane wave, or light from some point source, shining into some external front-end optics. That external optics could include lenses, optical fibers, or mode converters such as photonic lanterns or multiplane light converters in front of the integrated photonic mesh.

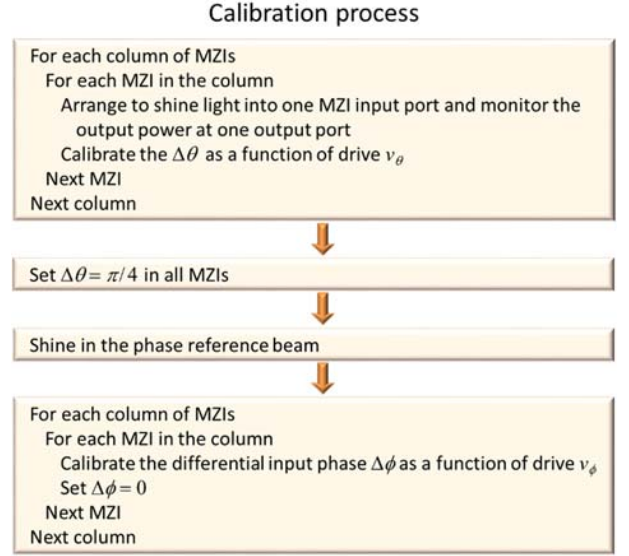

Fig. 4. Outline of the calibration process.

Integrated optical systems may hold a relatively stable phase calibration (especially if care is taken to make physical path lengths essentially equal for all beams that must interfere). In external optics, however, phase delays may well drift significantly with temperature or time. Furthermore, we may want to make changes to the front-end optics, such as changing focusing or even the modal decomposition basis of a complex system like a multiplane light converter. So, calibrating to that external phase reference also means we can "calibrate out" all the static phase behavior of such front-end optics, and can recalibrate for any changes in that phase behavior.

In operation, for each MZI we will have some drive $v_\theta$ (e.g., a voltage) that we are using to adjust $\Delta\theta$, and similarly some drive $v_\phi$ that we are using to adjust $\Delta\phi$. For example, $v_\theta$ might be the voltage used to drive a $\theta_P$ phase shifter, and similarly $v_\phi$ might be the drive voltage for a $\phi_T$ phase shifter. Calibrating means deducing the functions $\Delta\theta(v_\theta)$ and $\Delta\phi(v_\phi)$ (e.g., based on "look-up" tables of specific calibrated values) for each MZI. We use the functions $\Delta\theta(v_\theta)$ and $\Delta\phi(v_\phi)$ directly if we are running the mesh as a multimode analyzer so that we can deduce the $\Delta\theta$ and $\Delta\phi$ values for each MZI from the $v_\theta$ and $v_\phi$ values that have been set in the self-configuration. For running the mesh as a multimode generator, we can then numerically invert these calibration functions $\Delta\theta(v_\theta)$ and $\Delta\phi(v_\phi)$ to get functions $v_\theta(\Delta\theta)$ and $v_\phi(\Delta\phi)$ for each MZI; these tell us what drives $v_\theta$ and $v_\phi$ to apply to get the desired $\Delta\theta$ and $\Delta\phi$ in each MZI. We start with the calibration of the $\Delta\theta$ for each MZI and then move on to calibrating their $\Delta\phi$ behavior. We give detailed calibration procedures and formulas in the Supplement Section S7, but we can summarize the overall approach here (Fig. 4.)

The first part of the calibration is to calibrate the $\Delta\theta(v_\theta)$ function for each MZI. We do this by arranging optically that there is input power in just one of the two input ports (e.g., the "Top" port) of a given MZI. Then we use what we can call a "co-sinusoidal proportional calibration" to deduce $\Delta\theta(v_\theta)$ as we vary $v_\theta$; this is a relatively standard approach, though specifically this version is done using measured minimum and maximum output powers rather than presuming a minimum power of zero.

We can arrange for such an input power at just one port of an MZI in the mesh either in a "forwards" or "backwards" $\Delta\theta$ calibration approach. In a "forwards" calibration of the mesh, we require the input calibration power to be controllably in only one input port of the mesh at a time. Detectors for the calibration can be on the "drop ports" of the MZI, or we can use just the overall output power in the mesh output waveguide (though then the calibration of different MZIs cannot be "parallelized"). In a "backwards" calibration, we instead shine just one fixed backwards beam into the "output" of the mesh, but we need be able to monitor the power emerging backwards from the inputs to perform the calibration. In both schemes, we proceed through the MZIs, in a corresponding "forwards" or "backwards" sequence. (For a diagonal line self-configuring layer, because it is essentially symmetric from front to back, we can also run a "forwards" $\Delta\theta$ calibration with a single source, in a manner analogous to the "backwards" sequence here. See also [34].)

Having calibrated $\Delta\theta(v_\theta)$ for each MZI, to calibrate the $\Delta\phi(v_\phi)$ function for each MZI, essentially we set all the MZIs to $\Delta\theta = \pi/4$ (so the MZIs are each behaving like 50:50 beamsplitters overall). Then we shine in a "phase reference" beam over all the inputs; as discussed above, this phase reference can be one that is "in front of" any front-end optics. Using this reference, we then calibrate $\Delta\phi(v_\phi)$ for each MZI. By using the "co-sinusoidal proportional calibration" approach, importantly, we do not have to have equal input powers on the two ports of a given MZI; hence, for this calibration, essentially only the phase "shape" of that phase reference beam matters, not its intensity "shape". We work from the input MZIs progressively "forwards" through successive columns, starting from the "input" column; in each case, we set $\Delta\phi = 0$ in a given MZI once we have calibrated it.

This phase calibration approach, including setting all MZIs in "earlier" columns to $\Delta\phi = 0$ as we proceed to calibrate those in "later" columns, also means we automatically "calibrate out" any fixed phase delays in the system. Equivalently, it means that, for analysis or generation, we can proceed for each MZI as if

$$\phi_{av} = 0 \text{ if } \Delta\phi = 0 \text{ and } \theta_{av} = 0 \text{ if } \Delta\theta = 0 \quad (17)$$

Of course, if we change any of $\phi_T$, $\phi_L$, $\theta_P$ and/or $\theta_W$ in use, and we certainly will as we set $\Delta\theta$ and $\Delta\phi$ for given MZIs, we will change $\theta_{av}$ and $\phi_{av}$ by known amounts as a result, and we have to include that in our overall calculations, but we can start from the simple "baseline" as given by Eq. (17). For example, if we are using the $\phi_T$ and $\theta_P$ phase shifters to adjust the MZI, then, starting from this calibration and using Eqs. (1) - (5)

$$\Delta\theta = \theta_P, \ \Delta\phi = \phi_T, \ \phi_{Tot} = \phi_{av} + \theta_{av} + \frac{\pi}{2} = \frac{\phi_T}{2} + \frac{\theta_P}{2} + \frac{\pi}{2} \quad (18)$$

and we will use these relations when calculating the overall matrix M for a given MZI from Eq. (7).

There will obviously still be an additional overall phase delay in propagating through the system, but it will be the same for all paths, at least within additive multiples of $2\pi$, and hence does not matter for the purposes of multimode generation or analysis, where we only care about relative phases and amplitudes in the corresponding outputs or inputs.

Incidentally, if we want to calibrate a system with multiple self-configuring "layers" (see the Supplement Section S1), then we use a similar overall calibration process as we proceed to calibrate the "next" (and any subsequent) self-configuring layer(s). Specifically, we will set all blocks in preceding layers to $\Delta\phi = 0$ when we shine in the external phase reference. A later layer then treats all preceding layers as if they are just some extended "front-end" optics as we calibrate to that external phase reference.

One important point in both the calibration and operation of these networks is that we are presuming that cross-talk between the settings of the elements is negligible; otherwise, we are not able to adjust one element without unintentionally affecting another element. In the engineering of such systems, it will be important to minimize any such effects, such as thermal cross-talk [37].

Hence, we can have a straightforward and progressive calibration approach that could be completely automated. If we use the "backwards" $\Delta\theta(v_\theta)$ calibration, the calibration of the entire mesh can be accomplished with just two optical beams – one "backwards" power beam in the output waveguide for the $\Delta\theta(v_\theta)$ calibration, and one "forwards" phase reference beam for the $\Delta\phi(v_\phi)$ calibration.

## 7. Conclusions

We have shown that there is a simple, automatic method for analyzing the full amplitude and phase of the components of a multimode optical field and for generating any such field on a given modal basis. Effectively, in one automatic process we perform all the necessary interferometry in a mesh of two-beam interferometers. This process may be fast enough to measure such multimode fields in real time, e.g., on microsecond time-scales.

We also show how this system can be calibrated automatically. Importantly, all these processes require only simple sequential algorithms and arithmetic calculations, based physically on power minimization or maximization in photodetectors as we adjust each parameter one by one. Hence we can have a simple and fast method for full analysis and arbitrary generation of multimode fields in optics, for potential applications in communications, processing and sensing of various kinds.

**Funding**. Air Force Office of Scientific Research FA9550-17-1-0002.

**Acknowledgment**. I thank Sunil Pai for many stimulating discussions.

See the Supplement for supporting content.

# Analyzing and generating multimode optical fields using self-configuring networks: supplement

David A. B. Miller

*Ginzton Laboratory, Stanford University, 348 Via Pueblo Mall, Stanford CA 94305 USA*
*dabm@stanford.edu*

This document provides supplementary information to "Analyzing and generating multimode optical fields using self-configuring networks". Section S1 gives supporting definitions and detail on the topology of self-configuring layers. Section S2 gives a full analysis of general loss-less Mach-Zehnder interferometers (MZIs). Section S3 derives general properties of loss-less beamsplitters. Section S4 derives the necessary phase shifts in directional couplers used as beamsplitters. The transpose and phase-conjugating properties of unitary reciprocal networks are formally derived in Section S5. Section S6 gives the detailed algorithm and formulas for running a multimode generator. Explicit procedures for automatic calibration of the mesh are given in Section S7.

## S1 – Self-configuring network layers

Here we clarify and extend the topological discussion of self-configuring networks, including the ideas of "layers" and "columns". First, a definition, at least for our purposes, and as effectively defined previously [1,2]:

A self-configuring optical network is one that aligns itself to in an input optical field to perform a useful function on that field, based on simple feedback loops between a detector or detectors and controllable networks elements, such as phase shifters or possibly controllable couplers, without requiring any calculations [1,2].

Architectures and algorithms have been proposed for such self-configuring networks for what can be called "forward-only" or "feed-forward" networks [1-4], which are ones in which light flows only in one direction, from inputs to outputs, without topological loops or reflections [4] that would lead to light going backwards through the same network. If light can reflect or complete loops within the network, then we cannot apply simple progressive algorithms to configure them; changing an element "later" in the network can then lead to a change in the field "earlier" in the network, which means we may have to change the settings of the element "earlier" in the network again to compensate for that, thereby breaking a simple progression from "earlier" to "later" in setting up the networks.

Specifically, collections of universal 2x2 blocks, such as MZIs or controllable couplers and phase shifters [5] can form such forward-only networks. These are capable of performing arbitrary linear operations on mutually coherent light at a given frequency in a set of input waveguides when using sufficiently sophisticated architectures [2], including in architectures that can be fully self-configured automatically [1,2,6-12].

The simplest such self-configuring architectures form "self-aligning beam couplers" [1], which can take all the power in a set of (mutually coherent) inputs and route it to just one output, which we can call here the "signal" output (in the main text, because we do not need this distinction, we just call this the "output"). Such networks also have additional outputs, which can be called "drop-ports"; those may be used in monitoring and feedback loops in the self-aligning beam coupler functionality itself, though are not otherwise used as outputs in that application. Overall, including such drop-port outputs, there are generally equal total numbers of inputs and outputs in such a "self-aligning beam coupler" architecture. These nominally unused drop-port outputs in one such "self-aligning beam coupler" architecture can then be used as inputs for additional such units [1,2]. In that case, we can usefully think each such "self-aligning beam coupler" unit as a "self-configuring layer" in a possibly larger self-configuring system (See Fig. S1).

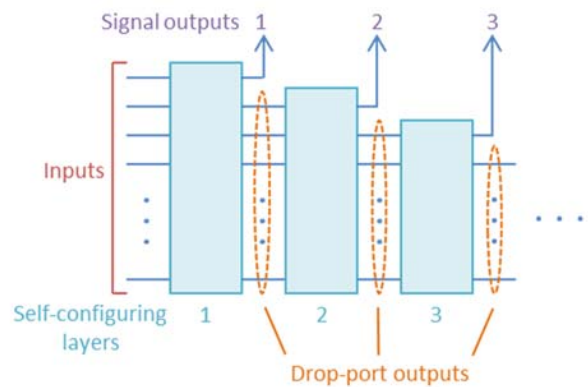

Fig. S1. Illustration of use of successive cascaded self-configuring layers (each of which can be considered to be a self-aligning beam coupler) to form a larger self-configuring network, with the drop-port outputs of one layer becoming the inputs of the next layer.

In such a case, any optical input that is mathematically orthogonal to the "configured" input for the first such layer is passed from these drop-port outputs to become the inputs of a subsequent self-configuring layer. In such a way, multiple such self-configuring layers can be cascaded to form a larger self-configuring architecture that is capable of separating multiple orthogonal input beams to the numbered outputs in Fig. S1
.

These cascaded architectures are introduced in [1,2]. These can be extended first to perform arbitrary unitary operations, and then can be extended also to implement full non-unitary matrix operations between inputs and outputs [2]. These full non-unitary matrix operations can be constructed based on the singular-value decomposition of the matrix, as introduced in [2]; that results in two unitary matrix operations, one at the input and one at the output, with a line of modulators connecting them. In the singular-value decomposition picture, we can consider the matrix being constructed using giving a set of orthogonal input "modes" (i.e., input vectors of amplitudes) that couple to a set of orthogonal output "modes" (i.e., output vectors of amplitudes).

These full non-unitary meshes may be self-configured by shining the desired input "modes" one by one into the input side of the mesh for "training" the input self-configuring layers one by one, and similarly training the output side of the mesh by shining in backwards (technically, phase-conjugated) versions of the corresponding desired output "modes". Because any optical device at a given wavelength can be described in terms of a singular value decomposition, this approach allows (and proves the possibility of) an arbitrary linear optical component at that wavelength, at least on this modal basis.

Two explicit architectures have been proposed for such self-configuring layers (or self-aligning beam couplers) – the "diagonal line" and the "binary tree", as shown in Fig. 1 in the main text, based on 2x2 blocks that can be implemented, for example, with MZIs.

For any given block in a layer, in general an input can be "external" – i.e., directly from an input to the layer, or "internal" – i.e., coming from the output of another block in the layer. Three distinct input configurations (Fig. S2) are used in the self-configuring layers discussed here: (a) a "binary tree" unit, which has two "internal" inputs; (b) a "diagonal line" unit, which has one "internal" input and one "external" input; and (c) an "input" unit, which has two "external" inputs. In the full binary tree mesh, the first column of 2x2 blocks are all "input" units, and all others are "binary tree". In the diagonal line mesh, the block in the first column is an "input" unit, and all the others are "diagonal line" units.

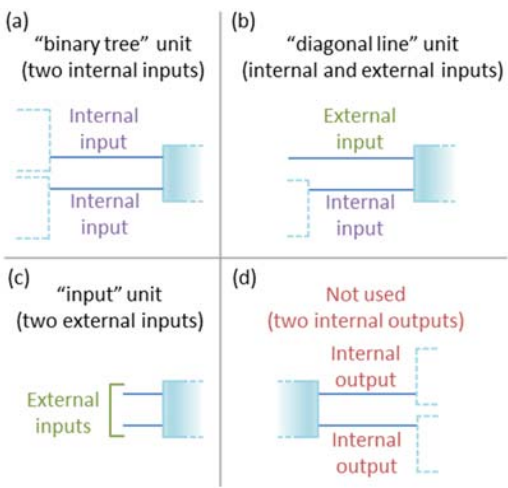

Fig. S2. Different configurations for connections to 2x2 blocks in a mesh architecture. (a), (b) and (c) in various combinations make up the self-configuring layers discussed here. Configuration (d) is not used in those, however; one block never connects its outputs to two other blocks in the same self-configuring layer.

Another configuration, Fig. S2(d), in which the two outputs of block are connected to inputs of other blocks within the same layer, does not occur within these self-configuring layers. This kind of configuration is common in architectures, such as the rectangular mesh [13], that cannot be decomposed into self-configuring layers, and that do not support overall self-configuration in the sense described here (though they can be configured with other algorithms – see, e.g., [4,12]).

Incidentally, generally these 2x2 blocks can be made from MZIs using any of allowed choices of 2 phase shifters discussed in the main text (Section 3) – i.e., any choice of two phase shifters as long as one is inside the MZI, though the "input" unit case of Fig. S2(c) also requires that there is at least one phase shifter on one of the inputs (so at least one of the $\phi_T$ or $\phi_L$ phase shifters in the notation of Fig. 3 of the main text.) These blocks can also be made using other approaches, such as controllable directional couplers [5].

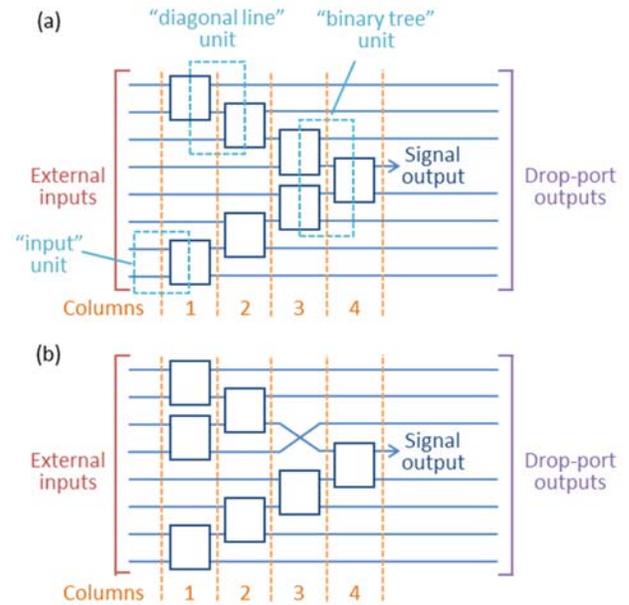

Fig. S3. Other example self-configuring layer architectures. (a) a "V" architecture with two diagonal lines joined with a final "binary tree" unit. (b) an architecture with a binary tree at the top and a diagonal line at the bottom. Blocks in the same column can be configured at the same time.

In addition to the simple (full) diagonal line and (full) binary tree self-configuring layers of Fig. 1 in the main text, other architectures of self-configuring layers are possible with different combinations of diagonal line units and binary tree units. Two examples are sketched in Fig. S3.

In addition to never using the "two internal outputs" configuration of Fig. S2(d), there are two other interesting conditions for constructing the self-configuring layers considered here.

1) It is desirable that there is only a single path from any input to the signal output. Otherwise, there is redundancy in the network; if there are two such paths, we would just have to arbitrarily choose to fix the settings on one such path, and configure the other as needed. (Using the "two internal outputs" configuration of Fig. S2(d) can lead to exactly such redundancies.)

2) Each input must have a possible connection to the signal output. (To be universal, the output must in general be a linear combination of all the inputs, hence this necessary condition.)

All the networks shown here (in Fig. 1 in the main text and Fig. S3) obey both of these criteria.

Of these various possible networks, the (full) binary tree is the shortest in terms of the number of blocks between the external inputs and the signal output. The (full) diagonal line has the characteristic that it is reversible; it can be run either with inputs on the left and outputs on the right or the other way round. In cascaded self-configuring layers, as

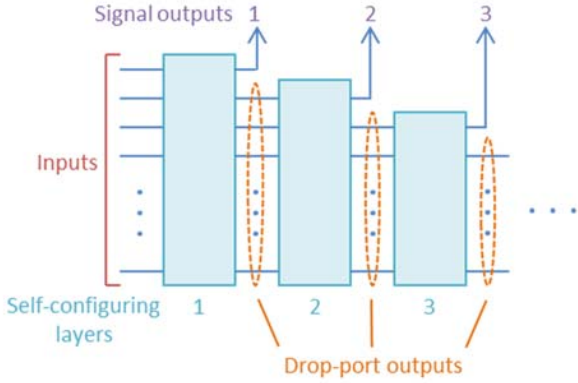

Fig. S1, it can lead to a shorter network overall than with binary-tree layers, and does not require crossing waveguides to extract the signal outputs from different layers.

## S2 – Analysis of a general lossless Mach-Zehnder interferometer

In analyzing the MZI of Fig. 3 in the main text, we work progressively through the MZI, considering the amplitudes and phases of the different paths and adding the results at the output ports. We take a monochromatic field of time-dependence $\exp(-i\omega t)$ in our analysis, which is consistent with a common approach of writing a "forward-propagating" wave as $\exp(i[kz - \omega t])$ in optics, and also with the common "physicist's" choice, as in Schrödinger's time-dependent wave equation [15]. A phase delay of $\phi$ radians then corresponds to multiplying by a factor $\exp(i\phi)$.

With our choice above that the beam passing from one waveguide to the other in the beamsplitters is delayed by 90°, which will hold for all directional couplers run in "first order" (whether or not they have a 50:50 beamsplitter ratio), the corresponding phase delay factor is

$$\exp(i\phi_{BS}) = \exp\left(i\frac{\pi}{2}\right) \equiv i \qquad (S1)$$

In considering a beamsplitter itself, we choose to label its ports in the same way as we have labelled the ports of the "cube" beamsplitter in Fig. 3(b) in the main text (so "Top", "Left", "Right" and "Bottom"). As in Section S3 below, we presume field "reflectivities" $r_{TR}$ ("Top" to "Right") and $r_{LB}$ ("Left" to "Bottom") field and similarly field "transmissivities" $t_{LR}$ and $t_{TB}$. We show in Section S3 that for a loss-less beamsplitter $|r_{TR}|^2 = |r_{LB}|^2 \equiv |r|^2$ (Eq. (S37)), and similarly $|t_{TB}|^2 = |t_{LR}|^2 = |t|^2 = 1 - |r|^2$ (Eq. (S38). Now we take the beamsplitter itself to have no phase shifts other than the 90° phase delay in crossing from one waveguide to the other (so, in "transmission" through the beamsplitter); all other phase shifts associated with propagation will be included instead in the phase delays $\phi_T$, $\phi_L$, $\theta_P$ and $\theta_W$ in the rest of the MZI.

With that choice, for a given beamsplitter $r$ can be taken to be real (and positive), and

$$t = \sqrt{1-r^2}\exp\left(i\frac{\pi}{2}\right) = i\sqrt{1-r^2} \qquad (S2)$$

Because in an MZI we have two beamsplitters which may have different field reflectivities and transmissivities, we need to label those as $r_1$ and $t_1$ for beamsplitter BS1 (on the "left" in Fig. 3(a) in the main text) and $r_2$ and $t_2$ for beamsplitter BS2 (on the "right" in Fig. 3(a) in the main text).

So, we continue our formal analysis, now for the full MZI of Fig. 3 in the main text with its "Top", "Left", "Right", and "Bottom" ports and with corresponding forward (complex) propagating mode amplitudes $a_T$, $a_L$, $a_R$, and $a_B$, respectively. Working progressively through the MZI, adding the amplitudes on various different paths, we have

$$\begin{aligned}a_R &= a_T \exp(i\phi_T)\{r_1 r_2 \exp(i\theta_P) + t_1 t_2 \exp(i\theta_W)\} \\ &+ a_L \exp(i\phi_L)\{r_1 t_2 \exp(i\theta_W) + t_1 r_2 \exp(i\theta_P)\}\end{aligned} \qquad (S3)$$

Taking out a common factor of $\exp(i\theta_{av})$ with $\theta_{av} = (\theta_P + \theta_W)/2$ (Eq. 1 in the main text), and using $\Delta\theta = \theta_P - \theta_W$ (Eq. 2 in the main text)

$$\begin{aligned}a_R &= a_T \exp\left[i(\phi_T + \theta_{av})\right]\left\{r_1 r_2 \exp\left(i\frac{\Delta\theta}{2}\right) + t_1 t_2 \exp\left(-i\frac{\Delta\theta}{2}\right)\right\} \\ &+ a_L \exp\left[i(\phi_L + \theta_{av})\right]\left\{r_1 t_2 \exp\left(-i\frac{\Delta\theta}{2}\right) + t_1 r_2 \exp\left(i\frac{\Delta\theta}{2}\right)\right\}\end{aligned}$$
(S4)

From Eq. 3 in the main text ($\phi_{av} = (\phi_T + \phi_L)/2$), Eq. 4 in the main text ($\Delta\phi = \phi_T - \phi_L$), and Eq. 5 in the main text ($\phi_{Tot} = \phi_{av} + \theta_{av} + \pi/2$) we have

$$a_R = -i\exp(i\phi_{Tot})$$
$$\times \begin{bmatrix} a_T \exp\left(i\frac{\Delta\phi}{2}\right)\left\{r_1 r_2 \exp\left(i\frac{\Delta\theta}{2}\right) + t_1 t_2 \exp\left(-i\frac{\Delta\theta}{2}\right)\right\} \\ + a_L \exp\left(-i\frac{\Delta\phi}{2}\right)\left\{r_1 t_2 \exp\left(-i\frac{\Delta\theta}{2}\right) + t_1 r_2 \exp\left(i\frac{\Delta\theta}{2}\right)\right\} \end{bmatrix}$$
(S5)

Similarly, by the symmetry of the MZI structure, we deduce the corresponding result for the "bottom" output amplitude by swapping "top" and "left" and "upper" and "lower" and correspondingly inverting the differential phase shifts in Eq. (S5), obtaining

$$a_B = -i\exp(i\phi_{Tot})$$
$$\times \begin{bmatrix} a_T \exp\left(i\frac{\Delta\phi}{2}\right)\left\{r_1 t_2 \exp\left(i\frac{\Delta\theta}{2}\right) + t_1 r_2 \exp\left(-i\frac{\Delta\theta}{2}\right)\right\} \\ + a_L \exp\left(-i\frac{\Delta\phi}{2}\right)\left\{r_1 r_2 \exp\left(-i\frac{\Delta\theta}{2}\right) + t_1 t_2 \exp\left(i\frac{\Delta\theta}{2}\right)\right\} \end{bmatrix}$$
(S6)

So, the most general form of the matrix $\mathsf{M}$ relating outputs to inputs for the MZI (as in Eq. 6 in the main text) becomes (Eq. (7) in the main text)

$$\mathsf{M} = \exp(i\phi_{Tot})\mathsf{M}_s(\Delta\theta)\mathsf{M}_\phi(\Delta\phi)$$

Here $\mathsf{M}_\phi(\Delta\phi)$, as in Eq. (8) in the main text, represents the effect of the difference between the input phase shifts on the "Top" and "Left" arms. The remaining effect of the MZI is then given by the matrix

$$\mathsf{M}_s(\Delta\theta) = -i\begin{bmatrix} \{r_1 r_2 e_+ + t_1 t_2 e_-\} & \{r_1 t_2 e_- + t_1 r_2 e_+\} \\ \{r_1 t_2 e_+ + t_1 r_2 e_-\} & \{r_1 r_2 e_- + t_1 t_2 e_+\} \end{bmatrix} \quad (S7)$$

where $e_+ = \exp(i\Delta\theta/2)$ and $e_- = \exp(-i\Delta\theta/2)$. When we consider the detailed calibration of the mesh (Supplement 1 Section S7), we need specific results for the "top-left" (i.e., first row and first column) matrix element $M_{11}$ and the "bottom-left" (i.e., second row and first column) element $M_{21}$. Using the facts that $r_1$ and $r_2$ are real (by choice) and $t_1$ and $t_2$ (as given correspondingly by Eq. (S2)) are purely "positive" imaginary, then

$$|M_{11}|^2 = a - b\cos\Delta\theta \quad (S8)$$

where $a$ and $b$ are positive numbers

$$a = 1 - r_1^2 - r_2^2 + 2r_1^2 r_2^2 \quad (S9)$$

$$b = 2r_1 r_2 \sqrt{1-r_1^2}\sqrt{1-r_2^2} \quad (S10)$$

Similarly,

$$|M_{21}|^2 = 1 - a + 2\cos\Delta\theta \quad (S11)$$

In the main text and in Section S7, we also need the $\mathsf{M}_\phi(\Delta\phi)$ matrix for the special case of 50:50 beamsplitters, i.e.,

$$r_1 = r_2 = 1/\sqrt{2} \quad (S12)$$

$$t_1 = t_2 = i/\sqrt{2} \quad (S13)$$

in which case $\mathsf{M}_s(\Delta\theta)$ becomes

$$\mathsf{M}_s(\Delta\theta) = \begin{bmatrix} \sin\left(\frac{\Delta\theta}{2}\right) & \cos\left(\frac{\Delta\theta}{2}\right) \\ \cos\left(\frac{\Delta\theta}{2}\right) & -\sin\left(\frac{\Delta\theta}{2}\right) \end{bmatrix} \quad (S14)$$

## S3 – Properties of a loss-less beamsplitter

Here we derive the formal properties of a loss-less beamsplitter. This analysis is similar to that of [16], expressed here in our notation, and with some added explicit algebra steps and results that we need. Consider a beamsplitter as in Fig. 3(b) in the main text. Now we presume the beamsplitter has field reflectivities $r_{TR}$ and $r_{LB}$ as well as field transmissivities $t_{LR}$ and $t_{TB}$. These reflectivities and transmissivities are complex numbers, and they are referred to the input and output "faces" or ports. So, for example, with some complex field amplitude $a_T$ in the "Top" beamsplitter port, the complex field amplitude at the "Right" beamsplitter port would be $r_{TR}a_T$, which would include all phase shifts between the top input face or port and the right output face or port. With similar definitions for corresponding terms for other ports, we could write the effect of the beamsplitter as a matrix equation

$$\begin{bmatrix} a_R \\ a_B \end{bmatrix} = \begin{bmatrix} r_{TR} & t_{LR} \\ t_{TB} & r_{LB} \end{bmatrix}\begin{bmatrix} a_T \\ a_L \end{bmatrix} \quad (S15)$$

or, explicitly,

$$a_R = r_{TR}a_T + t_{LR}a_L \quad (S16)$$

and

$$a_B = t_{TB}a_T + r_{LB}a_L \quad (S17)$$

Now we consider that the power in any beam is proportional to the modulus squared of the field amplitude, so, in appropriate units, the power in the beam at the "Right" beamsplitter port is

$$|a_R|^2 = |r_{TR}|^2|a_T|^2 + |t_{LR}|^2|a_L|^2 + t_{LR}r_{TR}^* a_L a_T^* + r_{TR}t_{LR}^* a_T a_L^* \quad (S18)$$

and the power in the beam at the "Bottom" beamsplitter port is

$$|a_B|^2 = |t_{TB}|^2|a_T|^2 + |r_{LB}|^2|a_L|^2 + r_{LB}t_{TB}^* a_L a_T^* + t_{TB}r_{LB}^* a_T a_L^* \quad (S19)$$

So the total output power is

$$|a_R|^2 + |a_B|^2 = \left(|t_{TB}|^2 + |r_{TR}|^2\right)|a_T|^2 + \left(|t_{LR}|^2 + |r_{LB}|^2\right)|a_L|^2 \\ + \left(t_{LR}r_{TR}^* + r_{LB}t_{TB}^*\right)a_L a_T^* + \left(t_{TB}r_{LB}^* + r_{TR}t_{LR}^*\right)a_T a_L^* \quad (S20)$$

Now we note that the last two terms in Eq. (S20) are complex conjugates of one another, so Eq. (S19) becomes

$$|a_R|^2 + |a_B|^2 = \left(|t_{TB}|^2 + |r_{TR}|^2\right)|a_T|^2 + \left(|t_{LR}|^2 + |r_{LB}|^2\right)|a_L|^2 \\ + 2\operatorname{Re}\left[\left(t_{LR}r_{TR}^* + r_{LB}t_{TB}^*\right)a_L a_T^*\right] \quad (S21)$$

Now suppose that we vary the relative phase of the two input beam amplitudes $a_L$ and $a_T$ while keeping their magnitudes constant. As a result, the phase of the term $(t_{LR}r_{TR}^* + r_{LB}t_{TB}^*)a_L a_T^*$ will change, which means the real part of this term will change. But none of the other terms in Eq. (S21) change, and, in particular, the left side of the equation will not change. Hence, since this equation must hold for arbitrary relative phases of the inputs, we can conclude that, in this presumed lossless beamsplitter, for conservation of power

$$t_{LR}r_{TR}^* + r_{LB}t_{TB}^* = 0 \quad (S22)$$

Hence

$$t_{LR}r_{TR}^* = -r_{LB}t_{TB}^* \quad (S23)$$

Writing

$$t_{LR} = |t_{LR}|\exp(i\phi_{LR}), \quad t_{TB} = |t_{TB}|\exp(i\phi_{TB}) \\ r_{TR} = |r_{TR}|\exp(i\phi_{TR}), \quad r_{LB} = |r_{LB}|\exp(i\phi_{LB}) \quad (S24)$$

so the various $\phi_{TR}$, $\phi_{LB}$, $\phi_{TB}$, and $\phi_{LR}$ correspond to phase delays in propagating through the system between their respective beamsplitter ports (presuming an $\exp(-i\omega t)$ time dependence in our analysis, as discussed in Section S2). Then, from Eq. (S23)

$$|t_{LR}||r_{TR}|\exp(i[\phi_{LR} - \phi_{TR}]) = -|r_{LB}||t_{TB}|\exp(i[\phi_{LB} - \phi_{TB}]) \quad (S25)$$

Taking the modulus of both sides of Eq. (S25) gives

$$|t_{LR}||r_{TR}| = |r_{LB}||t_{TB}| \quad (S26)$$

So, Eq. (S25) gives

$$\exp(i[\phi_{LR} - \phi_{TR}]) = -\exp(i[\phi_{LB} - \phi_{TB}]) \quad (S27)$$

Noting that

$$\exp(i[\pm \pi]) = -1 \quad (S28)$$

Eq. (S27) gives

$$\exp(i[\phi_{LR} + \phi_{TB} - \phi_{TR} - \phi_{LB}]) = \exp(i[\pm \pi]) \quad (S29)$$

So we conclude that

$$\phi_{LR} + \phi_{TB} - \phi_{TR} - \phi_{LB} = \pm \pi + 2n\pi \quad (S30)$$

for some integer $n$. (This is often stated with the additional $2n\pi$ presumed. For most practical beamsplitter designs, it may well be that $n = 0$, but we certainly could design beamsplitters in which it is not – for example, some very long directional coupler beamsplitter.)

Further obvious relations are obeyed by such a lossless beamsplitter. From Eq. (S21) we directly conclude that

$$|a_R|^2 + |a_B|^2 = (|t_{TB}|^2 + |r_{TR}|^2)|a_T|^2 + (|t_{LR}|^2 + |r_{LB}|^2)|a_L|^2 \quad (S31)$$

Choosing $a_L = 0$ and requiring power conservation in Eq. (S31) gives

$$|t_{TB}|^2 + |r_{TR}|^2 = 1 \quad (S32)$$

Similarly, choosing $a_T = 0$ gives

$$|t_{LR}|^2 + |r_{LB}|^2 = 1 \quad (S33)$$

Eq. (S26) leads to a simple conclusion that

$$\frac{|r_{TR}|}{|t_{TB}|} = \frac{|r_{LB}|}{|t_{LR}|} \quad (S34)$$

so the "split ratio" of this loss-less beamsplitter is the same whether we start from the "top" input port or the "bottom" input port.

Also, squaring both sides of Eq. (S26), and substituting using Eqs. (S32) and (S33) gives

$$(1 - |r_{LB}|^2)|r_{TR}|^2 = |r_{LB}|^2(1 - |r_{TR}|^2) \quad (S35)$$

So

$$|r_{TR}|^2 - |r_{LB}|^2|r_{TR}|^2 = |r_{LB}|^2 - |r_{LB}|^2|r_{TR}|^2 \quad (S36)$$

So, adding $|r_{LB}|^2|r_{TR}|^2$ to both sides

$$|r_{TR}|^2 = |r_{LB}|^2 \equiv |r|^2 \quad (S37)$$

Hence, the losslessness of the beamsplitter allows us to conclude that there is just one "power" reflectivity, which we can call $|r|^2$, for the beamsplitter, and similarly just one "power" transmissivity

$$|t_{TB}|^2 = |t_{LR}|^2 = |t|^2 = 1 - |r|^2 \quad (S38)$$

regardless of which input port we start with.

## S4 - Phase on "crossing over" in a directional coupler beamsplitter

Conservation of power in any lossless beamsplitter, as discussed in Section S3, leads to a general relation between the phase shifts between the various input and output ports [2,16]. If we use the notation of Section S3 (defining the total phase delays between the various beamsplitter ports T, L, B, and R, in the notation of Fig. 3(b) in the main text as $\phi_{TR}$, $\phi_{LB}$, $\phi_{TB}$, and $\phi_{LR}$), then conservation of power in a lossless beamsplitter requires [2,16] $\phi_{LR} + \phi_{TB} - \phi_{TR} - \phi_{LB} = \pm\pi + 2n\pi$ (Eq. (S30) of Section S3) for some integer $n$.

Suppose now that the beamsplitter is perfectly symmetric – so we could not tell any difference if we simultaneously exchanged the top and bottom and left and right ports. For the case of a cube beamsplitter, symmetry would also require that we had a perfect cube with the partially reflecting surface on the diagonal. For a directional coupler splitter, left-right physical reflection symmetry (i.e., reflecting in a plane perpendicular to the guides and half way along the length of the directional coupler) together with top-bottom reflection symmetry (i.e., reflecting along a line in the middle between the waveguides) would give such equivalent symmetry.

Such symmetry requires

$$\phi_{TR} = \phi_{LB} \quad (S39)$$

and

$$\phi_{TB} = \phi_{LR} \quad (S40)$$

So, from Eq. (S30) we deduce that

$$\phi_{TR} - \phi_{TB} = \pm\frac{\pi}{2} \quad (S41)$$

and

$$\phi_{LB} - \phi_{LR} = \pm\frac{\pi}{2} \quad (S42)$$

(within any additive integer multiple of $\pi$ on the right of each of Eqs. (S41) and (S42)). So, for any such symmetric beamsplitter, if we shine into only one input port, there is necessarily a $\pi/2$ (90°) degree phase difference (which could be a lead or a lag) between the two output ports.

(Note, incidentally, that an MZI run as a beamsplitter is not symmetric in this way. If an MZI has equal phase delays in both arms, it is in the "cross" state, which is equivalent to no "reflectivity", e.g., from the "top" to the "right" or from the "left" to the "bottom". To create any such "reflectivity", we need to unbalance the phase shifts in the two arms, which then means the actual MZI is not symmetric any more. Hence, an MZI operated as a beamsplitter does not have to obey Eqs. (S41) and (S42), and in fact it does not, which can be verified from the form of the matrix in Eq. (S7) of Section S2, for example. It does, however, still obey the more general condition on the relations between phases in a beamsplitter as in Eq. (S30) [2,16].)

The most common waveguide beamsplitter design for such mesh networks is a 50:50 directional coupler structure with coupling between two adjacent, identical waveguides. As discussed, for example, in [17] (Eqs. 3.21 and 3.22), the antisymmetric "second" coupled mode of two identical coupled waveguides propagates with a faster phase velocity than the symmetric "first" coupled mode. If we start out with the wave in one guide – say, the left guide – then the wave corresponds to an equal superposition of these first and second modes, adding in the left guide and cancelling in the right guide.

Now we can construct a physical argument, here for the case of a 50:50 beamsplitter, that resolves the sign of the phase change correctly. Now let these coupled modes propagate. There is then a point along the guides at which antisymmetric mode has advanced by 90° compared to the symmetric mode. At this point, as a result of this summation of two equal waves (the symmetric and antisymmetric mode components) that are now 90° out of phase, the phase of the net field in the left guide will therefore be 45° ahead of the symmetric mode phase, and the power in the left guide will be reduced by a factor of two; there will be an equal power in the right guide, thereby performing the 50:50 split of power. The antisymmetric mode component in the right guide started out in antiphase compared to symmetric mode component in the right guide, which could view as being 180 degrees "behind" the symmetric component. By this 50:50 split length, the antisymmetric mode component will only be 90° behind the symmetric mode component, and the phase of the net component in the right guide will be 45° behind the symmetric mode component. Therefore, the light in the right guide at this 50:50 split length will be 45° behind the phase of the symmetric mode component, and so will overall be 90° behind the light in the left guide. We can consider this to be a "first order" directional coupler 50:50 splitter since it has the shortest length at which we achieve such a 50:50 split. So the phase change in such a "first-order" directional coupler 50:50 splitter in "going from" the left guide to the right guide is an apparent 90°, and it is a delay. (It is, of course, possible to construct this argument algebraically, but we wanted to present an argument that was not reliant on sign conventions.)

Note also that the effective phase velocity in each guide will be half way between the phase velocities of the even and odd coupled modes. For weak coupling, these two phase velocities are approximately equally split round about the "single guide" phase velocity, so the net phase velocity in each guide will be the "single guide" phase velocity. Hence the light in each guide tend to propagate at the "single guide" phase velocity, with this effective 90° phase shift between the "straight through" and "crossing" paths in the coupler.

Note that, though we have constructed this argument for a 50:50 beamsplitter, in fact it applies for any beamsplitting ratio; note Eqs. (S41) and (S42) are derived independent of that ratio. In an actual directional coupler, this 90° phase delay continues as we increase the coupler length up to the point of total coupling from one guide to the other (which would then be a "0:100" beamsplitter). After that point, increasing the coupler length further causes the coupling to reduce, and the phase delay also changes sign, becoming now a phase lead of 90° (or a delay of 270°) until we reach the next length at which no net power couples over (so a "100:0" beamsplitter), and so on.

To complement this discussion, we can construct an algebraic argument. We presume two identical waveguides, 1 (e.g., the "left" one) and 2 (e.g., the "right" one), running parallel to one another in the $z$ direction, and that are weakly coupled. The "uncoupled" mode in each guide would have some spatial mode field amplitude for an appropriate polarization direction (e.g., transverse electric or transverse magnetic) that we could write, respectively as

$$E_{1u} = f(x_1, y_1) \text{ and } E_{2u} = f(x_2, y_2) \quad (S43)$$

for $x$ and $y$ coordinates centered correspondingly in each guide.

Now we consider the modes of the coupled system, in a typical "weak coupling" approach as in coupled mode theory [17], that considers just the $E_{1u}$ and $E_{2u}$ as being a sufficient basis for describing solutions to the coupled problem. (This is similar to other "coupled well" problems, e.g., as in a tight-binding approach in quantum mechanics [14].) In such weak-coupling models, the coupled eigenmodes of the system become symmetric (*S*) and antisymmetric (*A*) combinations that, for waves of angular frequency $\omega$, propagate in the $z$ direction with $k$-vector magnitudes $k_S$ and $k_A$. Based on our knowledge of such systems [17], $k_A < k_S$, corresponding to a faster phase velocity for the antisymmetric mode. So, we can write these modes, including their propagation in the $z$ direction, as

$$\begin{aligned} E_S &= (E_{1u} + E_{2u})\cos(k_S z - \omega t) \\ E_A &= (E_{1u} - E_{2u})\cos(k_A z - \omega t) \end{aligned} \quad (S44)$$

With this choice, a wave that starts out effectively as the "uncoupled" mode of guide 1 at $z = 0$ with no field in guide 2, can be written as

$$\begin{aligned} E &= E_S + E_A = (E_{1u} + E_{2u})\cos(k_S z - \omega t) \\ &+ (E_{1u} - E_{2u})\cos(k_A z - \omega t) \\ &= E_{1u}\left[\cos(k_S z - \omega t) + \cos(k_A z - \omega t)\right] \\ &+ E_{2u}\left[\cos(k_S z - \omega t) - \cos(k_A z - \omega t)\right] \\ &= 2E_{1u}\cos(k_{av} z - \omega t)\cos(\Delta k z) \\ &- 2E_{2u}\sin(k_{av} z - \omega t)\sin(\Delta k z) \end{aligned} \quad (S45)$$

where

$$k_{av} = \frac{k_S + k_A}{2} \text{ and } \Delta k = \frac{k_S - k_A}{2} \quad (S46)$$

Since $E_{1u}$ and $E_{2u}$ are the "uncoupled" mode forms in guides 1 and 2 respectively, then we can write, for the overall amplitude of the "mode" in guide 1

$$M_1 \propto \cos(k_{av} z - \omega t)\cos(\Delta k z) \quad (S47)$$

and, noting that $-\sin(\theta) = \cos(\theta + \pi/2)$, for the overall amplitude of the "mode" in guide 2

$$M_2 \propto \cos\left(k_{av} z + \frac{\pi}{2} - \omega t\right)\sin(\Delta k z) \quad (S48)$$

So, for $0 < \Delta k z < \pi/2$, in which range both $\cos(\Delta k z)$ and $\sin(\Delta k z)$ are positive, the amplitude $M_2$ in guide 2 is delayed in phase by $\pi/2$ compared to the amplitude $M_1$ in guide 1. This range $0 < z < \pi/2\Delta k$ encompasses the first complete range over which power couples progressively from starting entirely in guide

1 to ending entirely in guide 2, so over that whole range the wave in guide 2 is delayed by $\pi/2$. If we continue to larger $z$, so, for example in the range $\pi/2\Delta k < z < \pi/2\Delta k$, then $\cos(\Delta kz)$ becomes negative (though $\sin(\Delta kz)$ remains positive). The expression for $M_2$ can conveniently remain the same, but we can rewrite

$$M_1 \propto -\cos(k_{av}z - \omega t)|\cos(\Delta kz)| \\ = \cos(k_{av}z + \pi - \omega t)|\cos(\Delta kz)| \quad (S49)$$

So now the "mode" in guide 1 can be viewed as being delayed by $\pi/2$ compared to that in guide 2, inverting the situation compared to that for $0 < z < \pi/2\Delta k$. So, once the power starts to couple "back" from guide 2 to guide 1, the phase of the field in guide 1 can be viewed as being delayed by $\pi/2$ compared to that in guide 2. This situation reverses back and forward for successive such ranges as we go to successive "higher orders" in the directional coupler behavior.

## S5 – Transpose and phase-conjugating properties of unitary reciprocal networks

The properties we prove here are generally known and the proofs are straightforward, so much so that it may be difficult to find them stated explicitly in the literature. We give them because we need to use these results in different situations here, with a good understanding of where these results come from and just what the necessary assumptions are.

Suppose that we have an input vector of $N$ waveguide mode complex amplitudes

$$|\alpha\rangle \equiv \begin{bmatrix} \alpha_1 \\ \alpha_2 \\ \vdots \\ \alpha_N \end{bmatrix} \quad (S50)$$

and we have a linear reciprocal network represented by an $N \times N$ unitary matrix $\mathsf{U}$. Then the corresponding output vector of waveguide mode complex amplitudes is

$$|\beta\rangle \equiv \begin{bmatrix} \beta_1 \\ \beta_2 \\ \vdots \\ \beta_N \end{bmatrix} \quad (S51)$$

with

$$\mathsf{U}|\alpha\rangle = |\beta\rangle \quad (S52)$$

Specifically, then, looking at the relation between two elements of these vectors, we would have

$$\beta_m = u_{mn}\alpha_n \quad (S53)$$

where $u_{mn}$ is the "$mn$"th element ($m$th row, $n$th column) of the matrix $\mathsf{U}$. Now suppose that, instead of shining light in the "inputs" of the network, we instead shine a vector of amplitudes

$$|\gamma\rangle \equiv \begin{bmatrix} \gamma_1 \\ \gamma_2 \\ \vdots \\ \gamma_N \end{bmatrix} \quad (S54)$$

backwards into the "outputs" of the network, generating a vector of "backwards" amplitudes

$$|\eta\rangle \equiv \begin{bmatrix} \eta_1 \\ \eta_2 \\ \vdots \\ \eta_N \end{bmatrix} \quad (S55)$$

emerging from the "inputs" of the network. In this backwards direction, we write the matrix that relates these new "backwards" amplitudes to one another as $\mathsf{V}$, so by this definition we have

$$|\eta\rangle = \mathsf{V}|\gamma\rangle \quad (S56)$$

Now, by presumption, this network is reciprocal. Specifically, the coupling constant between specific input and output ports is the same in both directions, both in amplitude and in phase delay. So, specifically

$$\eta_n = u_{mn}\gamma_m \quad (S57)$$

This means that

$$\mathsf{V} = \mathsf{U}^T \quad (S58)$$

(the transpose of the original matrix $\mathsf{U}$). This result itself is useful and general: If we run a unitary network backwards, the corresponding matrix is the transpose of the "forward" one. We could call this the "transpose" property of unitary networks run in reverse.

Suppose, now, that the vector of backwards amplitudes shone into the "outputs" is the phase conjugate of $|\beta\rangle$, i.e., the vector

$$|\gamma\rangle = |\beta\rangle^* \equiv \begin{bmatrix} \beta_1^* \\ \beta_2^* \\ \vdots \\ \beta_N^* \end{bmatrix} \quad (S59)$$

[Note that, if a phase of a given output is "lagging" (retarded), then in the phase conjugate, the corresponding backwards phase-conjugate input will have a phase that is "leading" (advanced). In ordinary optics, then, if a beam is diverging as it propagates to the "right", then the phase conjugate beam will be converging as it propagates to the "left".] Then, using the definition of the conjugate transpose (i.e., Hermitian adjoint)

$$\mathsf{U}^\dagger \equiv (\mathsf{U}^T)^* \quad (S60)$$

together with the definition of unitarity

$$\mathsf{U}^\dagger \mathsf{U} = \mathsf{I} \quad (S61)$$

(where $\mathsf{I}$ is the $N \times N$ identity matrix), and Eqs. (S52), (S56), and (S58), we have

$$\mathsf{U}^T|\beta\rangle^* = (\mathsf{U}^\dagger|\beta\rangle)^* = (\mathsf{U}^\dagger \mathsf{U}|\alpha\rangle)^* = |\alpha\rangle^* \quad (S62)$$

So, if in the forward direction we have $\mathsf{U}|\alpha\rangle = |\beta\rangle$ (Eq. (S52)), then, if we shine the phase conjugate vector of amplitudes $|\beta\rangle^*$ backwards into the "outputs", the vector of backwards amplitude emerging from the "inputs" is $|\alpha\rangle^*$. We could call this the "phase-conjugating" property of reciprocal unitary networks.

## S6 – Settings for generating a desired backward beam

### Settings for a specific MZI block

We consider first how an MZI block must be configured if it takes all the power in some (forward) input vector $[a_T \ a_L]^T$ and routes it to one (forward) output. Quite generally, because the matrix $\mathsf{M}$ describing a MZI block is unitary (so $\mathsf{M}^\dagger \mathsf{M} = \mathsf{I}$, the identity matrix), then Eq. (6) in the main text leads to

$$\mathsf{M}^\dagger \begin{bmatrix} a_R \\ a_B \end{bmatrix} = \mathsf{M}^\dagger \mathsf{M} \begin{bmatrix} a_T \\ a_L \end{bmatrix} = \begin{bmatrix} a_T \\ a_L \end{bmatrix} \quad (S63)$$

where $\mathsf{M}^\dagger$ is the Hermitian adjoint (i.e., complex conjugate of the transpose) of $\mathsf{M}$. Then from Eq. (7) in the main text

$$\mathsf{M}^\dagger(\phi_{Tot}, \Delta\theta, \Delta\phi) = \exp(-i\phi_{Tot})\mathsf{M}_\phi^\dagger(\Delta\phi)\mathsf{M}_s^\dagger(\Delta\theta) \quad (S64)$$

(Note that, as usual, we have inverted the order of the matrix multiplications in taking the Hermitian adjoint of a matrix product. Note also that for 50:50 beamsplitters, $\mathsf{M}_s$ is actually Hermitian – so equal to its own Hermitian adjoint – though that is not generally the case for other beamsplitter ratios.) Explicitly, using Eq. (8) in the main text for $\mathsf{M}_\phi$, and presuming 50:50 beamsplitters, so using Eq. (9) in the main text for $\mathsf{M}_s$, we have

$$\mathsf{M}^\dagger(\phi_{Tot}, \Delta\theta, \Delta\phi) = \exp(-i\phi_{Tot})$$
$$\times \begin{bmatrix} \exp\left(-i\dfrac{\Delta\phi}{2}\right) & 0 \\ 0 & \exp\left(i\dfrac{\Delta\phi}{2}\right) \end{bmatrix} \begin{bmatrix} \sin\left(\dfrac{\Delta\theta}{2}\right) & \cos\left(\dfrac{\Delta\theta}{2}\right) \\ \cos\left(\dfrac{\Delta\theta}{2}\right) & -\sin\left(\dfrac{\Delta\theta}{2}\right) \end{bmatrix} \quad (S65)$$

So, for routing all this input power to the "Right" MZI optical output, we have

$$\begin{bmatrix} a_T \\ a_L \end{bmatrix} = \mathsf{M}^\dagger \begin{bmatrix} a_R \\ 0 \end{bmatrix} \quad (S66)$$

which gives

$$\begin{bmatrix} a_T \\ a_L \end{bmatrix} = a_R \exp(-i\phi_{Tot}) \begin{bmatrix} \exp\left(-i\dfrac{\Delta\phi}{2}\right)\sin\left(\dfrac{\Delta\theta}{2}\right) \\ \exp\left(i\dfrac{\Delta\phi}{2}\right)\cos\left(\dfrac{\Delta\theta}{2}\right) \end{bmatrix} \quad (S67)$$

(Note, incidentally, that we are still physically imagining that we are running the system forwards; if $a_R$ is the output we get at the "Right" MZI output and we are getting zero output at the "Bottom" MZI output, then this is what the input amplitudes $a_T$ and $a_L$ must have been.)
Writing

$$\begin{bmatrix} a_T \\ a_L \end{bmatrix} \equiv \begin{bmatrix} |a_T|\exp(i\psi_T) \\ |a_L|\exp(i\psi_L) \end{bmatrix} \quad (S68)$$

where $\psi_T$ and $\psi_L$ are real (and $\psi_T$ and $\psi_L$ correspond to whatever are the phase delays on these respective inputs), then, using Eq. (S67)

$$\frac{a_T}{a_L} = \frac{|a_T|}{|a_L|}\exp(i[\psi_T - \psi_L]) = \exp(-i\Delta\phi)\tan\left(\frac{\Delta\theta}{2}\right) \quad (S69)$$

So we should set

$$\Delta\phi = \psi_L - \psi_T \quad (S70)$$

$$\Delta\theta = 2\arctan\left(\frac{|a_T|}{|a_L|}\right) \quad (S71)$$

Note that, since $|a_T|$ and $|a_L|$ are necessarily positive, Eq. (S71) will return an answer for $\Delta\theta$ that lies in the range 0 to $\pi$. In this approach, then, we will always be choosing $\theta_P$ in the range $\theta_L$ to $\theta_L + \pi$, so the phase delay in the upper guide will always be chosen equal to or larger than that in the lower guide.

If instead we consider an MZI block in which we are routing all the power to the "bottom" optical output, so

$$\begin{bmatrix} a_T \\ a_L \end{bmatrix} = \mathsf{M}^\dagger \begin{bmatrix} 0 \\ a_B \end{bmatrix} \quad (S72)$$

then

$$\begin{bmatrix} a_T \\ a_L \end{bmatrix} = a_B \exp(-i\phi_{Tot}) \begin{bmatrix} \exp\left(-i\dfrac{\Delta\phi}{2}\right)\cos\left(\dfrac{\Delta\theta}{2}\right) \\ -\exp\left(i\dfrac{\Delta\phi}{2}\right)\sin\left(\dfrac{\Delta\theta}{2}\right) \end{bmatrix} \quad (S73)$$

So, now

$$\frac{a_T}{a_L} = \frac{|a_T|}{|a_L|}\exp(i[\psi_T - \psi_L]) = -\exp(-i\Delta\phi)\cot\left(\frac{\Delta\theta}{2}\right)$$
$$= \exp(-i[\Delta\phi \pm \pi])\cot\left(\frac{\Delta\theta}{2}\right) \quad (S74)$$

So we should set

$$\Delta\phi = \psi_L - \psi_T \pm \pi \quad (S75)$$

$$\Delta\theta = 2\,\mathrm{arccot}\left(\frac{|a_T|}{|a_L|}\right) \quad (S76)$$

Note here that we will choose the + or − sign in Eq. (S75) as needed in practice to keep $\Delta\phi$ within whatever $2\pi$ range of phase delays we choose for it (e.g., 0 to $2\pi$ or possibly $-\pi$ to $\pi$). In this approach, again because $|a_T|$ and $|a_L|$ are necessarily positive, Eq. (S76) will result in $\Delta\theta$ in the range 0 to $\pi$, and so the phase delay in the upper guide will always be chosen equal to or larger than that in the lower guide. An alternative approach for this case of routing all the power to the "Bottom" output is to make a mirror image, reflecting in a horizontal plane, of the MZI configuration used for the "Right" output case. Then we end up putting the larger phase delay in the lower MZI arm (so $\Delta\theta$ is in the range 0 to $-\pi$) and we avoid the additional $\pm\pi$ of Eq. (S75).

**Settings for the entire mesh**

Now that we have deduced how a given MZI must have been set so that it would have routed any particular pair of (forward) input complex amplitudes to one or other output, we can proceed to calculate the settings for the entire mesh to allow us to generate an arbitrary vector of (backward) amplitudes.

We start, then, by choosing the desired vector $|d\rangle$ to be generated as backward amplitudes from the "input" waveguides, and we calculate the vector $|c\rangle = (|d\rangle)^*$. We presume there are $Q$ columns in the mesh, numbered from the left-most column as column 1 as in Fig. 1 in the main text. Each column will have a corresponding vector $|c^{(q)}\rangle$ of (mathematical) "forwards" amplitudes. So now the algorithm is as follows.

Use $|c\rangle$ as the column input vector $|c^{(1)}\rangle \equiv |c\rangle$ for the first (i.e., left-most) column

For each column $q$ (from 1 to $Q$)

For each block in the column

Using the appropriate elements of $|c^{(q)}\rangle$ to give the "top" ($a_T$) and "left" ($a_L$) amplitudes for each block in the column calculate the settings $\Delta\phi$ and $\Delta\theta$, from Eqs. (S70) and (S71) if the block optical output is the "right" port, or from Eqs. (S75) and (S76) if the block optical output is the "bottom" port, and physically apply those settings to the block.

Using the resulting matrix M for the block and the block's known hypothetical "forward" input amplitudes $a_T$ and $a_L$

calculate the (complex) amplitude ($a_R$ or $a_B$ as appropriate) in the optical output port using Eq. (6) in the main text.

Use these amplitudes to construct the corresponding vector $|c^{(q+1)}\rangle$ of optical output amplitudes from the column. (For a binary tree mesh, the vector $|c^{(q+1)}\rangle$ will have half as many elements as the vector $|c^{(q)}\rangle$, and for the diagonal line mesh it will have one fewer element.)

Next block

Next column

Once we have run this algorithm and applied all the resulting calculated settings to the mesh elements, the mesh is set so that shining light backwards into the "output" port of the mesh will lead to vector $|d\rangle$ of complex (relative) amplitudes and phases emerging backwards from the mesh "input" ports.

## S7 – Calibration

Here we give detailed expressions and algorithms for performing the $\Delta\theta$ and $\Delta\phi$ calibrations for the MZIs in the mesh, supporting the summary discussion in Section 6 of the main text. The $\Delta\phi$ calibration also takes care of "calibrating out" any differing fixed phase delays on the different paths inside the mesh as well as the fixed overall phase behavior of any "front-end" optics.

**$\Delta\theta$ calibration**

*Forwards and backwards approaches*

$\Delta\theta$ is the parameter that controls the "split ratio" of the MZI. For this calibration, we need an overall approach that allows us to arrange for power input in only one port of the MZI during its individual $\Delta\theta$ calibration. There are at least two ways that we can arrange that – "forward" calibration or "backward" calibration.

In forward calibration, we could arrange to illuminate only one input waveguide in the mesh at a time. That obviously allows us to put input power into only one port of each MZI in the input column in the binary tree mesh, and into each separate MZI in entire mesh for the diagonal line mesh. For successive columns of the binary tree, it will automatically be the case that, with only one such input power in the input column, any given MZI in a subsequent column can only have power in at most one input; so, with an appropriately chosen input in the first column, it is straightforward to arrange for power in only one input for an MZI in any other column.

In backwards calibration, for example of the binary tree mesh, we shine power backwards into the output port, so we only need one simple optical input power for this part of the calibration. We then calibrate the MZIs backwards. In this case, we need to be able to monitor the output power emerging backwards from the input ports. That could be arranged with "sampling" detectors for measuring "backwards" power in those "input" ports, or by observing the backwards output power from each such "input" port on a camera. Having calibrated the single MZI in the rightmost column based on detected power backwards at the "inputs", we can set that MZI to route power to a specific input of the MZI in a preceding column, and proceed accordingly backwards through the whole mesh. We can proceed similarly for the diagonal line with power backwards into the output. For the diagonal line mesh, because it is essentially symmetric "front" to "back", we can also run a scheme similar to the "backwards" calibration approach but in a forwards direction using instead a single beam in the "lowest" input (e.g., waveguide 8 in Fig. 1 in the main text). (This kind of calibration process for a diagonal line mesh is also discussed in [12], where the architecture is described as an "optical setup machine" and is used to provide input vectors for configuring some other mesh.)

Whether backwards or forwards calibration is better depends on whether it is practically easier to control input power separate to each mesh input or to monitor backwards power at each mesh input.

*Calibrating $\Delta\theta$ for a specific MZI*

For an MZI as in Fig. 3 in the main text, we presume, then, that we send an optical power input into just one input port (so, "Top" or "Left"), and that we can monitor the power in one output port (so, "Right" or "Bottom"). We have two different situations in a binary tree mesh as in Fig. 1 in the main text. In some MZIs there, such as K11, K13, and K21, the detector is on the "Right" output of the MZI (in the notation of Fig. 3 in the main text), and in others, such as K12, K14, K22, and K31, the detector is on the "Bottom" output. In the way we have chosen to draw the diagonal line architecture, with the diagonal line running from bottom-left to top-right, for all the blocks K1 to K7, the detector is on the "Bottom" output; we could also choose to draw the diagonal line architecture as going from top-left to bottom right, in which case all the detectors would be on the "Right" outputs of the blocks.

We also presume we know the sign of the phase delay induced by the drive. For example, in a silicon waveguide in which phase delay is introduced by heating, the phase delay increases with increasing voltage, current or power drive because the thermo-optic coefficient (rate of change of refractive index with temperature) is positive when operating at infrared wavelengths where silicon is nominally transparent.

For the sake of definiteness for the moment, we presume we use the "Top" port for the input and we analyze first the case where the output power monitoring detector is on the "Right" port.

In such a case, then, for input power $P_T$ in the "Top" input, taking the power as the modulus squared of the amplitude (in appropriate units), the power at the "Right" output is, from Eq. (6) in the main text

$$P_R = P_T \left|M_{11}\right|^2 \quad (S77)$$

where $M_{11}$ is the top-left (i.e., first row, first column) element of the matrix $\mathbf{M}$.

For 50:50 beamsplitters, from Eq. (9) in the main text

$$\left|M_{11}\right|^2 = \sin^2\left(\frac{\Delta\theta}{2}\right) = \frac{1}{2}[1-\cos\Delta\theta] \quad (S78)$$

More generally, even if the beamsplitters do not have a 50:50 split ratio, $\left|M_{11}\right|^2$ still has the general form (Eq. (S8), Section S2)

$$\left|M_{11}\right|^2 = \frac{P_R}{P_T} = a - b\cos\Delta\theta$$

where the positive numbers $a$ and $b$ are given in Section S2 in Eqs. (S9) and (S10), respectively. The expression $a - b\cos(\Delta\theta)$ always lies between 0 and 1, and Eq. (S8) shows too that even for arbitrary beamsplitter ratios (i.e., ones that are not necessarily "50:50"), this quantity is minimum at $\Delta\theta = 0$ and maximum at $\Delta\theta = \pi$.

In practice beamsplitters will never be perfectly 50:50, which can mean the output power may well never get exactly to zero, no matter what phase shifter settings are chosen. As a result, it is better to use Eq. (S8) than Eq. (S78) as the basis for calibrating $\Delta\theta$ because Eq. (S8) will give a much better calibration especially near to whatever is the actual minimum output power.

So, using Eqs. (S77) and (S8)

$$\cos\Delta\theta = \frac{1}{b}\left(a - \frac{P_R}{P_T}\right) \quad (S79)$$

Now, from Eq. (S8), the maximum and minimum possible values of $P_R$ for a given input power $P_T$ are, respectively

$$P_{Rmax} = (a+b)P_T \quad (S80)$$

$$P_{Rmin} = (a-b)P_T \quad (S81)$$

Using Eqs. (S80) and (S81) to construct substitutes for $a$ and $b$ in Eq. (S79) leads to

$$\cos\Delta\theta = \frac{P_{Rmax} + P_{Rmin} - 2P_R}{P_{Rmax} - P_{Rmin}} \quad (S82)$$

So to calibrate the relative phase shift $\Delta\theta$, with a given fixed input power $P_T$ only in the "Top" input, we should adjust the corresponding phase shift drive value $v_\theta$ over a full range to find and measure the maximum and minimum output powers $P_{Rmax}$ and $P_{Rmin}$. Then, for each of a desired set of drive values $v_\theta$, we would measure the output power $P_R(v_\theta)$ and deduce the corresponding $\Delta\theta(v_\theta)$ for each such value using

$$\Delta\theta(v_\theta) = \arccos\left(\frac{P_{Rmax} + P_{Rmin} - 2P_R(v_\theta)}{P_{Rmax} - P_{Rmin}}\right) \quad (S83)$$

thereby establishing a "look-up table" of pairs of drive and result values. (Above, we have called this kind of calibration approach "co-sinusoidal proportional calibration".) Possibly we would establish $\Delta\theta(v_\theta)$ as a smooth continuous function by an appropriate interpolation formula on this look-up table. We could then establish the inverse function $v_\theta(\Delta\theta)$ mathematically from $\Delta\theta(v_\theta)$, possibly also using interpolation to establish a corresponding smooth inverse function.

Note that the standard defined range of the inverse cosine (i.e., arccos), as used in Eq. (S83), is 0 to $\pi$, which is a suitable defined range for our purposes, and is sufficient to move from a "cross" state at $\Delta\theta = 0$ to a "bar" state at $\Delta\theta = \pi$.

Incidentally, because such a calibration approach uses only ratios of measured powers, it does not require that we calibrate the absolute power sensitivity of the detector; we only require that the detector gives an output that is linearly proportional to the input power and that it gives a "zero" output signal for zero input power (or that we subtract off any apparent output signal for zero actual input power).

For the case with the detector on the "Bottom" output, we have, analogously to Eq. (S77), for the "Bottom" power $P_B$

$$P_B = P_T \left|M_{21}\right|^2 \quad (S84)$$

with the general form (Eq. (S11), Section S2)

$$\left|M_{21}\right|^2 = 1 - a + 2\cos\Delta\theta$$

and then, analogously to Eq. (S82)

$$\cos\Delta\theta = \frac{2P_B - P_{Bmin} - P_{Bmax}}{P_{Bmax} - P_{Bmin}} \quad (S85)$$

We can then proceed analogously to the "Right" output case to calibrate $\Delta\theta(v_\theta)$.

The $\Delta\theta$ gives the same results if we perform the calibration backwards, with input powers on the right side of the MZI and outputs on the left. Then we just interchange "Top" for "Right" and "Bottom" for "Left" (and *vice versa*) in the above discussion.

### $\Delta\phi$ calibration

*Overall algorithm*

The $\Delta\phi$ calibration should be done after the $\Delta\theta$ calibration because, when calibrating $\Delta\phi$ in the scheme proposed here, we need to set a given MZI in the $\Delta\theta = \pi/4$ state (which means the entire MZI is effectively behaving as a 50:50 splitter). This $\Delta\phi$ calibration is also only correct if the MZI beamsplitters are themselves quite accurately 50:50 splitters; otherwise there will generally be a phase error in the estimate of the condition for $\Delta\phi = 0$.

As discussed in Section 6 of the main text, in calibrating phase we need to choose a phase reference, and that has to be optically defined by us. There is no intrinsic meaning to saying that light at two different points in space (such as at the "entrance" to two different waveguides) is "in phase"; we have to choose what we mean by that. So, we shine in some phase reference beam over all the inputs. That beam could be just a plane wave over the waveguide inputs or, more generally, could be some plane wave or distant point source illuminating whatever is the "front-end" optics of the entire system, or even some other phase front, such as a spherically expanding or focusing beam.

No matter what we use as our phase reference, as a result of shining in this phase reference beam, some power arrives at all the input waveguides, and we use that input field to calibrate $\Delta\phi$ for all the MZIs. Importantly, in the scheme we describe here, the incident powers do not have to be the same on the different waveguides.

This $\Delta\phi$ calibration is done in the "forward" direction. It is simplest if we have drop port detectors on each MZI, but if not, we should set all subsequent MZIs to route the power from the output port of the MZI of interest to an output detector for this calibration.

The phase calibration process is as follows, all with the phase reference shining into all inputs. We start with the MZI(s) in the first, "left-most" column. Such an MZI has direct input of the phase reference, without passing through any other MZIs. (Incidentally, this "column by column" process works for both the binary tree and diagonal line meshes.)

(i) We set the MZIs in this column to have $\Delta\theta = \pi/2$

(ii) Then we calibrate each MZI in this column using a "co-sinusoidal proportional calibration", similar to that used for the $\Delta\theta$ calibration above. (Explicit algebra is given below.) In this case, calibrating between the measured minimum and maximum output powers (instead of presuming the minimum power is zero) allows us to have unequal input powers and still perform the calibration. A simple such co-sinusoidal proportional calibration will first give us a calibration of $\Delta\phi$ over a range from 0 to $\pi$. We then need to extend that to a calibration over 0 to $2\pi$ (or alternatively $-\pi$ to $+\pi$ if we are driving a phase shifter pair differentially), by running the calibration over a second co-sinusoidal "cycle". In this way, we will establish appropriate calibration functions $\Delta\phi(v_\phi)$ and $v_\phi(\Delta\phi)$ (similarly to the $\Delta\theta(v_\theta)$ and $v_\theta(\Delta\theta)$ calibration functions) for each MZI.

(iii) We repeat this for all MZIs in a column

(iv) We then set all of these MZIs in the column to $\Delta\phi = 0$ (still retaining the $\Delta\theta = \pi/4$ setting).

We then repeat this process for the next column of MZIs, and so on for successive columns.

In this way, we calibrate $\Delta\phi$ for every MZI in the mesh, and all relative to our defined incident phase reference beam.

Now we can formally derive the algebra for the $\Delta\phi$ calibration of a given MZI.

*Calibrating $\Delta\phi$ for an MZI*

We presume that, with the phase reference beam shining over all the mesh inputs, we are calibrating an MZI using field amplitudes, $a_T$ for the mode in the "Top" single-mode guide input, and $a_L$ correspondingly in the "Left" input guide, and we take both of these to be real and positive, consistent with them having the same phase by definition.

An important practical point (as already mentioned) is that, for this calibration, $a_T$ and $a_L$ do not have to have the same magnitude, so the phase reference beam need not be of uniform intensity over the mesh inputs.

For this calibration, we presume we have already calibrated $\Delta\theta$ for the MZI, and we now specifically set $\Delta\theta = \pi/2$. Also, to get the calibration correct, we do need to presume that indeed the beamsplitters inside the MZI have an accurate 50:50 split ratio. In general, it would be possible to calibrate $\Delta\phi$ if we had other split ratios, but to get the $\Delta\phi = 0$ point correct, we would need to know the exact split ratios; different split ratios in general "shift" the interference minima and/or maxima away from the true $\Delta\phi = 0$ point, as can be verified by a detailed analysis based on the full form of M, as given in Section S2. (If we have some doubt about the fabricated split ratios, we could take the approach of [3,11], which allows us to construct effective 50:50 split ratios for beamsplitters in an MZI even with fabricated split ratios that are not 50:50.)

Presuming, then, $\Delta\theta = \pi/2$ and 50:50 beamsplitters, from Eq. (9) in the main text

$$\mathsf{M}_s(\Delta\theta) = \begin{bmatrix} 1/\sqrt{2} & 1/\sqrt{2} \\ 1/\sqrt{2} & -1/\sqrt{2} \end{bmatrix} \quad (S86)$$

(which, incidentally, means the MZI is set to function as if it were overall a 50:50 beamsplitter). So, with Eqs. (6) – (9) in the main text, for the amplitudes $a_R$ and $a_B$ in the "Right" and "Bottom" outputs, respectively,

$$\begin{bmatrix} a_R \\ a_B \end{bmatrix} = \exp(i\phi_{Tot}) \begin{bmatrix} 1/\sqrt{2} & 1/\sqrt{2} \\ 1/\sqrt{2} & -1/\sqrt{2} \end{bmatrix} \begin{bmatrix} \exp\left(i\dfrac{\Delta\phi}{2}\right) a_T \\ \exp\left(-i\dfrac{\Delta\phi}{2}\right) a_L \end{bmatrix}$$

$$= \frac{1}{\sqrt{2}} \exp(i\phi_{Tot}) \begin{bmatrix} \exp\left(i\dfrac{\Delta\phi}{2}\right) a_T + \exp\left(-i\dfrac{\Delta\phi}{2}\right) a_L \\ \exp\left(i\dfrac{\Delta\phi}{2}\right) a_T - \exp\left(-i\dfrac{\Delta\phi}{2}\right) a_L \end{bmatrix}$$
(S87)

Using the fact that $a_T$ and $a_L$ are both real, the incident powers are $P_T \equiv |a_T|^2 = a_T^2$ and $P_L \equiv |a_L|^2 = a_L^2$ at the "Top" and "Left" ports, respectively. So, for the powers $P_R \equiv |a_R|^2$ and $P_B \equiv |a_B|^2$ at the "Right" and "Bottom" ports respectively, using the fact that $a_T$ and $a_L$ are both also positive, we have

$$P_R = \frac{1}{2}\left[ P_T + P_L + \frac{\sqrt{P_T P_L}}{2}\left(\exp(i\Delta\phi) + \exp(-i\Delta\phi)\right)\right]$$
$$= \frac{P_T + P_L}{2} + \sqrt{P_T P_L}\cos(\Delta\phi) \quad (S88)$$

and similarly

$$P_B = \frac{1}{2}\left[ P_T + P_L + \frac{\sqrt{P_T P_L}}{2}\left(-\exp(i\Delta\phi) - \exp(-i\Delta\phi)\right)\right]$$
$$= \frac{P_T + P_L}{2} - \sqrt{P_T P_L}\cos(\Delta\phi) \quad (S89)$$

Now we can take a "co-sinusoidal proportional calibration" approach, similar to that used for calibrating $\Delta\theta$ above, in which we start out by scanning $\Delta\phi$ over a full range to establish and measure minimum and maximum powers first. Then, with some set of drives (e.g., voltages) $v_\phi$ for the $\Delta\phi$ phase shift, we calibrate the corresponding values $\Delta\phi(v_\phi)$ from the resulting cosinusoidal variation of the output power in one or other output port, establishing a "look-up table" of calibrated values.

Explicitly, rewriting expressions Eq. (S88) and (S89) in terms of the corresponding measured minimum and maximum powers $P_{Rmin}$, $P_{Bmin}$, $P_{Rmax}$, and $P_{Bmax}$, with

$$P_{Rmin} = P_{Bmin} = \frac{P_T + P_L}{2} - \sqrt{P_T P_L} \quad \text{(S90)}$$

$$P_{Rmax} = P_{Bmax} = \frac{P_T + P_L}{2} + \sqrt{P_T P_L} \quad \text{(S91)}$$

we can rearrange Eqs. (S88) and (S89) respectively to give, based on measuring $P_R$ (so, with a detector in the "Right" port), we deduce the corresponding value of $\Delta\phi$ from

$$\Delta\phi = \arccos\left(\frac{2P_R - P_{Rmax} - P_{Rmin}}{P_{Rmax} - P_{Rmin}}\right) \quad \text{(S92)}$$

or, based on measuring $P_B$ (so with a detector in the "Bottom" port),

$$\Delta\phi = \arccos\left(\frac{P_{Bmax} + P_{Bmin} - 2P_B}{P_{Bmax} - P_{Bmin}}\right) \quad \text{(S93)}$$

These expressions Eqs. (S92) and (S93) will always return results in the range from 0 to $\pi$. We can use them directly, therefore, to calibrate $\Delta\phi(v_\phi)$ in this range. One subtlety, however, is that we will want a total range of $2\pi$ for $\Delta\phi$ to make the MZI universal for our purposes. If we have phase shifters on both input arms, i.e., $\phi_T$ and $\phi_L$, and we drive them differentially, we might want this range to go overall from $-\pi$ to $+\pi$. In that case, we would calibrate with Eqs. (S92) and (S93) for "positive" differential drive, and with -1 times these expressions for "negative" differential drive. If we are running only with one phase shifter, and presuming that any phase shifters only give increasing phase delays with increasing drive, then for a phase shifter on only the "Top" arm (so $\phi_T$), we would use these expressions for the range 0 to $\pi$, and $2\pi$ minus these expressions as we continued up to a total of $2\pi$ of phase shift. If we only had a phase shifter on the "Left" arm (so $\phi_L$), then we would take the same approach, but all the phase shifts would be negative, so the negative of the approach for the "$\phi_T$-only" case, and we would run over a range of $-2\pi$ to zero for $\Delta\phi$.

So, we can summarize this $\Delta\phi$ calibration process:

1) Having already calibrated the "split ratio" for an MZI, we now set it for 50:50 split ratio (i.e., we set $\Delta\theta = \pi/2$).

2) We shine a phase reference beam into the two ("top" and "left") inputs, which defines what we mean by $\Delta\phi = 0$

3) Using the monitor detector on the "right" R (or "bottom" B) port as appropriate to the MZI, we scan $\Delta\phi$ over at least a range of $\pi$ to find maximum and minimum powers $P_{Rmax}$ and $P_{Rmin}$ (or $P_{Bmax}$ and $P_{Bmin}$).

4) Now as a function of drive $v_\phi$, for some set of suitably closely spaced values of $v_\phi$, we measure $P_R$ (or $P_B$), allowing us to calculate $\rho_{R\phi}(v_\phi)$ (or $\rho_{B\phi}(v_\phi)$), and hence deduce $\Delta\phi(v_\phi)$ using Eq. (S92) (or Eq. (S93)), and extending over a $2\pi$ range as discussed above.

5) By interpolation on the resulting "look-up table", we therefore can establish a suitable smooth function $\Delta\phi(v_\phi)$, and also invert it to establish a corresponding smooth function $v_\phi(\Delta\phi)$.